\documentclass[aps, prb,showpacs, amsfonts,amsmath,amssymb,superscriptaddress,twocolumn, 10pt]{revtex4-2}

\usepackage[utf8]{inputenc}
\usepackage[T1]{fontenc}
\usepackage[english]{babel}
\usepackage[colorlinks = true,linkcolor = blue,urlcolor= blue,citecolor = blue,anchorcolor = blue]{hyperref}
\usepackage{graphicx}
\usepackage{braket}
\usepackage{physics}
\usepackage{multirow}
\usepackage{booktabs}
\usepackage{orcidlink}
\usepackage{cellspace}
\def\doubleunderline#1{\underline{\underline{#1}}}

\begin{document}
\title{Effective Hamiltonians for Ge/Si core/shell nanowires from higher order perturbation theory}

\author{Sebastian Miles\orcidlink{0009-0005-6425-8072}}
\affiliation{QuTech and Kavli Institute of Nanoscience, Delft University of Technology, P.O. Box 4056, 2600 GA Delft, The Netherlands}

\author{A. Mert Bozkurt\orcidlink{0000-0003-0593-6062}}
\affiliation{QuTech and Kavli Institute of Nanoscience, Delft University of Technology, P.O. Box 4056, 2600 GA Delft, The Netherlands}

\author{Dániel Varjas\orcidlink{0000-0002-3283-6182}}
\affiliation{IFW Dresden and W\"urzburg-Dresden Cluster of Excellence ct.qmat, Helmholtzstrasse 20, 01069 Dresden, Germany}
\affiliation{Max Planck Institute for the Physics of Complex Systems,
N\"othnitzer Strasse 38, 01187 Dresden, Germany}
\affiliation{Department of Theoretical Physics, Institute of Physics,
Budapest University of Technology and Economics, M\H{u}egyetem rkp. 3., 1111 Budapest, Hungary}

\author{Michael Wimmer\orcidlink{0000-0001-6654-2310}}
\affiliation{QuTech and Kavli Institute of Nanoscience, Delft University of Technology, P.O. Box 4056, 2600 GA Delft, The Netherlands}

\begin{abstract}
We theoretically explore the electronic structure of holes in cylindrical Germanium/Silicon core/shell nanowires using a perturbation theory approach.
The approach yields a set of interpretable and transferable effective low-energy models for the lowest few sub-bands up to fifth order for experimentally relevant growth directions.
In particular, we are able to resolve higher order cross terms e.g., the dependency of the effective mass on the magnetic field.
Our study reveals orbital inversions of the lowest sub-bands for low-symmetry growth directions, leading to significant changes of the lower order effective coefficients.
We demonstrate a reduction of the direct Rashba spin-orbit interaction due to competing symmetry effects for low-symmetry growth directions.
Finally, we find that the effective mass of the confined holes can diverge yielding quasi flat bands interesting for correlated states.
We show how one can tune the effective mass of a single spin band allowing one to tune the effective mass selectively to its divergent points.

\end{abstract}
\maketitle

\section{Introduction}

Germanium (Ge) has played an important role in modern semiconductor development. 
In addition to the prospect of leveraging established silicon (Si) semiconductor infrastructure~\cite{zwerver_qubits_2022, scappucci_quantum-ready_2022}, Ge's unique material properties make it attractive for quantum technologies~\cite{scappucci_germanium_2021}.
Ge has been explored for diverse applications, including lasers~\cite{hauge_single-crystalline_2017, peeters_direct_2024, van_tilburg_stimulated_2024}, semiconductor-superconductor hybrids~\cite{tosato_hard_2023, valentini_parity-conserving_2024, aggarwal_enhancement_2021, lakic_quantum_2025, hinderling_direct_2024, kate_finite_2025}, and hole-based qubits~\cite{unseld_2d_2023, watzinger_germanium_2018,john_two-dimensional_2025,hofmann_assessing_2019, ungerer_charge-sensing_2023, van_riggelen-doelman_coherent_2024, stehouwer_exploiting_2025, hu_hole_2012}.
Isotopic purification enables Ge to achieve a low intrinsic nuclear-spin environment~\cite{prechtel_decoupling_2016}, which suppresses spin decoherence~\cite{bulaev_spin_2005}.
Additionally, its relatively low disorder allows for long mean free paths~\cite{vigneau_germanium_2019, xiang_gesi_2006}
Alloying Ge with other materials allows tailoring of specific properties.
For example, alloying with Si allows tuning the strain environment of Ge quantum wells~\cite{shimura_compressively_2024, lodari_lightly_2022, lodari_low_2021, martinez_variability_2022, costa_reducing_2024}.
Alloying with Sn modifies spin-orbit interactions~\cite{vecchio_fully_2025}. 
In Ge, the relevant hole degrees of freedom are p-type, which, combined with strong intrinsic spin-orbit coupling, lead to the formation of light- and heavy-hole bands~\cite{scappucci_germanium_2021, luttinger_motion_1955, baldereschi_spherical_1973}.
Confinement breaks this degeneracy and couples light- and heavy-holes.
Thus, confinement can provide advantages for e.g. gate implementation~\cite{bosco_phase-driving_2023, bosco_squeezed_2021} and enhanced noise resilience~\cite{svastits_readout_2025}.

A system combining strain engineering and confinement is the Ge/Si core/shell nanowire~\cite{conesa-boj_boosting_2017, goldthorpe_inhibiting_2009, goldthorpe_synthesis_2008, nguyen_diameter-independent_2014, nguyen_growth_2005}.
Such nanowires have been used to demonstrate a broad range of effects e.g., multiple Andreev reflection~\cite{ridderbos_multiple_2019}, quantum dots~\cite{brauns_anisotropic_2016}, and effects of superconductivity~\cite{ridderbos_hard_2020, wu_magnetic_2024}.
However, the geometry of nanowires introduces an additional degree of freedom: lattice orientation~\cite{conesa-boj_boosting_2017}.
In the regime where the radius is only a few orders of magnitude larger, we will show that different growth directions become a relevant effect that renormalize the spectrum.
Capturing all geometric effects does not require knowledge of the full Hamiltonian.
Instead, a few bands suffice to capture low-energy physics near the Fermi level.
These bands can be described using Schrieffer–Wolff perturbation theory~\cite{schrieffer_relation_1966, lowdin_note_1951, bravyi_schriefferwolff_2011, araya_day_pymablock_2025}.
This approach perturbatively decouples the low-energy orbital subspace from the rest of the Hilbert space.
In Ge, this approach has revealed phenomena such as the direct Rashba spin–orbit interaction~\cite{kloeffel_strong_2011, kloeffel_direct_2018, adelsberger_enhanced_2022}.
While successful for lower order effects, higher-order effects such as e.g., mutual influences of fields or lattice alignment have remained challenging within this approach.

In this work, we study both effects of lattice alignment and mutual field effects by performing higher order perturbation theory using the package Pymablock~\cite{araya_day_pymablock_2025}.
For each nanowire growth direction, we generate effective models expressed as polynomial expansions in applied perturbations.
We focus on perturbations by magnetic and electric fields, strain deviations, and longitudinal momentum.
The resulting effective Hamiltonians are easily interpretable and reveal the interdependencies and magnitudes of various perturbations.
The remainder of this paper is structured as follows.
Section~\ref{sec:model_pt} introduces the Hamiltonian and the perturbations considered, along with our strategy for generating effective Hamiltonians via Schrieffer–Wolff perturbation theory.
Section~\ref{sec:orbital_inversion} discusses how growth direction affects the low-energy orbitals.
Section~\ref{sec:g_fac} examines effective g-factors, while Section~\ref{sec:rashba} focuses on linear and cubic spin–orbit interaction coefficients.
Section~\ref{Sec:effective_mass} analyzes the tunability of the effective mass under external fields.
Finally, Section~\ref{sec:conclusion} summarizes our findings and offers an outlook.

\section{Model and perturbation theory} \label{sec:model_pt}

\subsection{Hamiltonian}
We model the valence hole bands in the Ge core of the nanowire with the Luttinger-Kohn Hamiltonian~\cite{luttinger_quantum_1956, bulaev_spin_2005}
\begin{align} \label{eq:h_lk_full}
\begin{split}
    H_{LK}
= \frac{\hbar^2}{2m}\bigg[
&\left(\gamma_1+\frac{5}{2}\gamma_2\right) k^2 \\
&- 2\gamma_2 \sum_i k_i^2 J_i^2 \\
&- 2\gamma_3 \sum_{i\neq j} \{k_i, k_j\}\{J_i, J_j\}
\bigg].
\end{split}
\end{align}
where $i,j = x,y,z$,  $k$ is the momentum, $J_i$ is a set of spin-$3/2$ matrices, $m$ is the mass of the valence band holes, and $\{A,B\}=AB+BA$ is the anticommutator. 
The parameters $\gamma_i$ are the material-specific Luttinger parameters.
We can split eq.~\eqref{eq:h_lk_full} into~\cite{baldereschi_cubic_1974, baldereschi_spherical_1973, kloeffel_strong_2011, kloeffel_direct_2018},
\begin{align} \label{eq:h_lk_split_symbolic}
    H_{LK} &= H_{\odot} + H_{k_z} + H_c.
\end{align}
The first term,
\begin{align} \label{LK:odot}
    H_{\odot}&=\frac{\hbar^2}{2m}\bigg[\left(\gamma_1+\frac{5}{2}\gamma_s\right)(k_x^2+k_y^2)\\ \nonumber
    &-2\gamma_s\left(k_x^2J_x^2+k_y^2J_y^2+\{k_x,k_y\}\{J_x,J_y\}\right)\bigg],
\end{align}
where $\gamma_1=13.35$ and $\gamma_s=(3\gamma_3+2\gamma_2)/5=5.11$ are the corresponding Luttinger parameters~\cite{winkler_spinorbit_2003}, captures the cylindrical invariant terms within the wire cross section.
Additional terms involving momenta parallel to the wire axis are collected in 
\begin{equation}\label{LK:LKz}
\begin{split}
        H_{k_z}&=\frac{\hbar}{2m}\bigg[\left(\gamma_1+\frac{5}{2}\gamma_s\right)k_z^2 +k_z^2J_z^2 \\
    &-2\gamma_s\big(\{k_y,k_z\}\{J_y,J_z\}+\{k_z,k_x\}\{J_z,J_x\}\big)\bigg]. 
\end{split}
\end{equation}
The sum of the first two terms of eq.~\eqref{eq:h_lk_split_symbolic} is spherically symmetric.
All parts of eq.~\eqref{eq:h_lk_full} that capture the cubic $O_h$ symmetry of the crystal lattice are collected in
\begin{align}\label{eq:cubic_term}
    H_c=\frac{\hbar^2\delta_\gamma}{2m}\left[\sum_i \left[k_i^2\left(2J_i^2-\frac{3}{2}\right)\right]-\frac{4}{5}(\vec{k}\cdot\vec{J})^2\right],
\end{align}
where $\delta_\gamma=\gamma_3-\gamma_2=1.44$ for Ge~\cite{winkler_spinorbit_2003}.

\begin{figure}
    \centering
    \includegraphics[width=\linewidth]{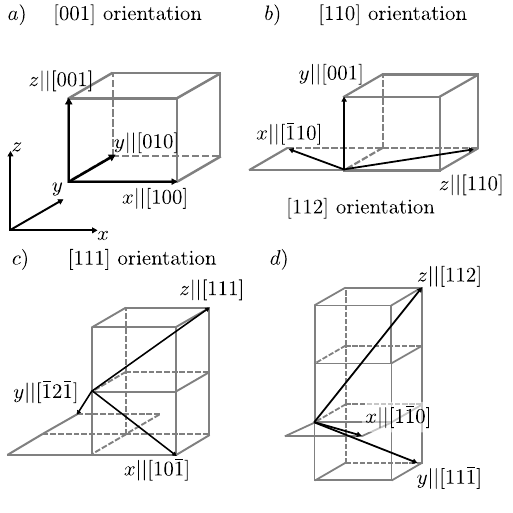}
    \caption{Illustration of the crystallographic directions and their bases considered in this work. By aligning the bases with high symmetry directions of the crystal, the effective $g^*$-tensor (c.f. Sec.~\ref{sec:g_fac}) and SOI tensor (c.f. Sec.~\ref{sec:rashba}) become purely diagonal and off-diagonal respectively (cf. App.~\ref{app_symmetry}). Only in $[112]$ the basis cannot fully be aligned with symmetry directions leading to cross terms between orthogonal directions.}
    \label{fig:crystal_orientations}
\end{figure}

The specific geometry of the nanowire is modeled by adding the cylindrically symmetric confinement potential
\begin{align}
    V(x,y)=\begin{cases}
        0 & r=\sqrt{x^2+y^2}\leq R_c \\
        \infty & r>R_c
    \end{cases},
\end{align}
where $R_c$ is the radius of the core of the wire, to the Hamiltonian, eq.~\eqref{eq:h_lk_split_symbolic}.

\begin{figure}
    \centering
    \includegraphics[width=\linewidth]{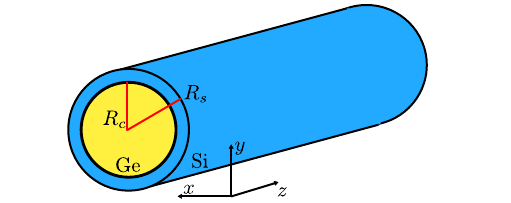}
    \caption{Schematic of the Ge/Si core/shell nanowire. The relevant parameters for strain (c.f. eq.~\eqref{LK:BP}) are the core- and shell-radius, $R_c$ and $R_s$ respectively. They tune an aggregated parameter we call $\delta_\epsilon$.}
    \label{fig:device}
\end{figure}

This approach assumes that the weight of the wavefunction in the shell of the wire is negligible, applicable to the low-energy states well localized in the core of the nanowire.
In the absence of the cubic term, $H_c$, the cylindrical symmetry of the confinement potential yields an additional conservation of angular momentum around the wire axis. 
Specifically, $H_{\odot}$ and confinement potential commute with the total angular momentum operator $F_z=L_z+J_z$, i.e. $[H_{\odot}+V(x,y), F_z]=0$ where $L_z=-i\partial_\varphi$~\cite{kloeffel_strong_2011}.

Due to the mismatch of the lattice constants, the Ge core of the wire is strained by the surrounding Si shell.
This alters the electronic properties of the holes.
Following~\cite{kloeffel_strong_2011, kloeffel_acoustic_2014, kloeffel_direct_2018} we model the strain exerted by the shell using the Bir-Pikus Hamiltonian~\cite{bir_symmetry_1974} in its lowest-order isotropic approximation,  reading
\begin{align} \label{LK:BP}
    H_{BP}&=b\delta_\epsilon J_z^2,
\end{align}
where $\delta_\epsilon \propto (R_s-R_c)/R_c$ is a parameter that encodes the isotropic strain exerted on the core.
This approximation allows us to retain the most dominant effect of strain, the renormalization of the energy gaps in the spectrum, while omitting additional anisotropy effects of the g-tensor~\cite{terrazos_theory_2021} from the details of the strain profile.

In experiments, the electronic properties can be tuned through electric and magnetic fields.
Magnetic effects are described by linear and cubic Zeeman terms,
\begin{align}
    H_Z&=2\kappa\mu_B \vec{B}\cdot\vec{J} \\
    H_{Z,3} &= 2q\mu_B \vec{B} \cdot (J_x^3, J_y^3, J_z^3)^T,
\end{align}
where $\kappa=3.41$ and $q=0.067$ are material coefficients of Ge~\cite{winkler_spinorbit_2003}.
Magnetic fields also induce an orbital effect.
This is modeled by replacing bare momentum operators with kinetic momentum operators.
In minimal coupling, this is $k\rightarrow k+\frac{e}{\hbar}\vec{A}(\vec{r})$, where $\vec{A}(\vec{r})$ is the corresponding vector potential.
We use the Landau gauge for $x$ and $y$ components of the vector potential and the symmetric gauge for the $z$ component:
\begin{align}\label{LK:gauge}
    \vec{A}=(-B_z y/2, B_z x/2, B_x y - B_y x)^T.
\end{align}
We collect all terms of $H_{LK}$ proportional to the magnetic field of the kinetic momentum in a Hamiltonian that we will refer to as $H_{LK, orb}$.
Electric field effects are included through the term:
\begin{align}\label{LK:He}
    H_E=-e\vec{r}\cdot \vec{E},
\end{align}
where the minus sign accounts for the charge of the holes. 

\subsection{Perturbation theory}

Our goal is to derive an effective description of the lowest hole orbital, referred to as the low-energy subspace.
This requires block diagonalization of the Hamiltonian.
We employ Schrieffer–Wolff perturbation theory~\cite{lowdin_note_1951, bravyi_schriefferwolff_2011} implemented in the Pymablock Python package~\cite{araya_day_pymablock_2025}.
We decompose the Hamiltonian as
\begin{align} \label{eq:h_pert_symb}
    H=H_0+H^\prime(\lambda).
\end{align}
$H_0$ is the sum of Hamiltonians with energy scales comparable to the excitation gap between ground and excited orbitals.
Specifically, $H_0$ contains at least $H_\odot$ and we add $H_c, H_{BP}$ in later sections. 
$H^\prime(\lambda)$ comprises all remaining terms treated perturbatively, parameterized by $\lambda\in{B_x,B_y,B_z,E_x,E_y,\delta_\epsilon,\delta_\gamma,k_z}$.
We discretize the in-plane coordinates and momenta using Kwant~\cite{groth_kwant_2014} while retaining the perturbative symbols $\lambda_i$ as variables.
We then numerically solve the stationary Schrödinger equation,
\begin{align} \label{eq:schroed_states}
H_0|\psi_{n,\sigma}\rangle = E_{n,\sigma} |\psi_{n,\sigma}\rangle,
\end{align}
where $n$ is the orbital quantum number and $\sigma$ is the effective spin of the specific level.
We extract the low-energy eigenstates $\psi_{1,\sigma}$ as the basis for the perturbation theory.

The representation of the resulting effective Hamiltonian strongly depends on the gauge of the low-energy eigenstates.
To simplify this representation, we exploit the symmetries of the Hamiltonian.
Throughout this work, $H_0$ is chosen to be time-reversal symmetric, such that the solutions of Eq.~\eqref{eq:schroed_states} are spin-degenerate.
Hence, we can choose the low-energy eigenstates to be Kramers partners, $\ket{\psi_{0,\downarrow}} = \mathcal{T} \ket{\psi_{0,\uparrow}}$.
We align the spin quantization axis along the nanowire axis and diagonalize the projected $J_z$ operator.
This allows a representation to satisfy $\bra{\psi_{0,i}}J_z\ket{\psi_{0,j}} \propto (\sigma_z){ij}$.
The relative phase of the basis states is then fixed so that, under the wire’s rotational symmetry, the low-energy subspace transforms as an effective spin-$1/2$.
This is achieved by choosing the gauge such that $\bra{\psi{0,\uparrow}}(J_x+iJ_y)\ket{\psi_{0,\downarrow}}$ is real and positive, where $J_x$ is perpendicular to the wire axis and parallel to a twofold rotation axis.
We consider the four different growth directions: $[001]$, $[110]$, $[111]$, and $[112]$~\cite{conesa-boj_boosting_2017}.
The corresponding $x$ axes are $[100]$ for $[001]$, $[\overline{1}00]$ for $[110]$, $[10\overline{1}]$ for $[111]$, and $[1\overline{1}0]$ for $[112]$ wires.
In Fig.~\ref{fig:crystal_orientations} we show the directions of the unit vector with respect to the cubic subgroups of the crystal symmetries.

Projecting the excited bands onto the two bands of the ground state orbital, we can represent the effective Hamiltonian in terms of a single set of Pauli matrices $\sigma_i$.
Each Pauli matrix is accompanied by a polynomial prefactor depending on the parameters $\lambda$ and their perturbative coefficients describing the magnitude of the effect.
For this particular case, we can write the effective model as
\begin{equation} \label{eqn:Heff2}
\begin{split}
    H_{\mathrm{eff}}&=\frac{\hbar^2}{2 m^*}k_z^2+\frac{\mu_B}{2} (\vec{B}\cdot \doubleunderline{g}^*)\cdot\vec{\sigma}+k_z \doubleunderline{\alpha_{L}}\cdot \vec{\sigma} \\
    &+\hbar^2 k_z^2 \frac{\partial}{\partial \vec{B}}\left[\frac{1}{m^*}\right]\vec{B}\cdot\vec{\sigma}+k_z^3 \doubleunderline{\alpha_{C}}\cdot \vec{\sigma}
\end{split}
\end{equation}
where $\vec{\sigma}$ is the vector of Pauli matrices, and the derivative $\partial_{\vec{B}}$ acts only on $1/m^*$.
Eq.~\eqref{eqn:Heff2} contains $\doubleunderline{g^*}$, the effective g-tensor, which we discuss in Sec. \ref{sec:g_fac}, $\doubleunderline{\alpha_{L}}$ and $\doubleunderline{\alpha_{C}}$, the linear and cubic spin-orbit tensors, discussed in Sec. \ref{sec:rashba}, as well as $m^*$ and $\partial_{\vec{B}}[1/m^*]$, the effective mass and its magnetic field dependent correction, discussed in Sec. \ref{Sec:effective_mass}.
To determine the content of each coefficient, we calculate polynomial expansions of our effective Hamiltonians as explained in App.~\ref{app:operator_transform}.
The specific effective coefficients can then be determined as polynomial prefactors of specific terms of that polynomial expansion.

To treat wires of all radii on equal footing, we introduce a characteristic set of scales.
We rescale energies by the characteristic level spacing $E_0=\hbar^2/m_eR_c^2$.
Furthermore, we define $\tilde{k}_z=R_ck_z$ and $\tilde{r}=r/R_c$.
The magnetic field, we define $\tilde{B}_i=eR_c^2B_i/\hbar$, the magnetic field per flux quantum through the wire cross section.
The electric field we change as $\tilde{E}_i=eR_cE_i/E_0$, the potential difference across the wire per level spacing.
Lastly, we rescale the strain as $\tilde{\delta}_\epsilon=\delta_\epsilon/E_0$.

\section{Growth direction and orbital inversion} \label{sec:orbital_inversion}

\begin{figure}
    \centering
    \includegraphics[width=\columnwidth]{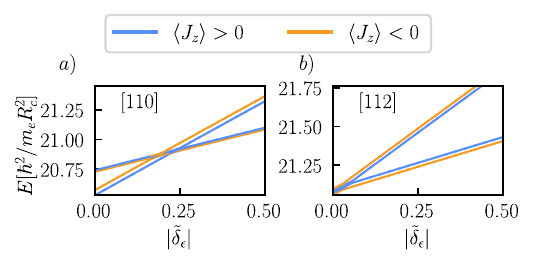}
    \caption{Exact diagonalization of the full Luttinger-Kohn Hamiltonian in the presence of a magnetic field $\vec{B}||x$ for the $[110]$ and $[112]$ growth directions. Reducing the symmetry from cylindrical to cubic causes orbital inversions in the $[110]$ direction and interlaced orbital states in the $[112]$ direction. In the former case, a perturbative treatment based on the lowest two eigenstates of the isotropic model is not applicable. In the latter case, the interlacing of orbitals prevents the definition of a well-defined effective g-factor.}
    \label{fig:orb_inv}
\end{figure}

\begin{figure}
    \centering
    \includegraphics[width=\linewidth]{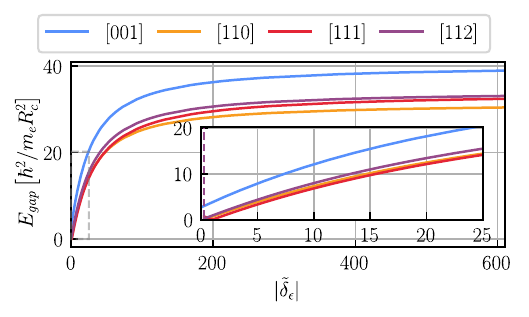}
    \caption{Excitation gap between the lowest two orbitals. In the low strain regime, the orbital inversion due to the cubic symmetry terms leads to a reduction of the excitation gap to up to $50\%$ for crystal orientations other than $[001]$.}
    \label{fig:energy_gap}
\end{figure}

To ensure the validity of the perturbation theory, we need to understand its limitations. 
The relevant scale for the perturbation theory is the excitation gap between ground and excited orbitals.
This gap becomes small in the limit of small strains.
The excitation gap must be compared with the renormalization effects introduced by $H_c$, eq.~\eqref{eq:cubic_term}.
This effect is most pronounced in regimes of low strain or small wire radius.
Specifically, we find an orbital inversion between ground and excited orbitals compared to the eigenstates of $H_{\odot}$~\cite{kloeffel_strong_2011, kloeffel_direct_2018}.
Fig.~\ref{fig:orb_inv} shows exact diagonalization results for the two lowest orbitals under a small spin-resolving magnetic field.
For the $[110]$ growth direction, orbital inversion occurs at finite strain, corresponding to approximately 1\% lattice mismatch for a core radius of 5 nm.
In contrast, for the $[112]$ direction, the orbitals are interlaced even without strain.
This observation is significant since effects such as DRSOI depend intimately on the size of the excitation gap~\cite{kloeffel_strong_2011, kloeffel_direct_2018}.
Near the orbital inversions, a single orbital perturbation theory becomes inapplicable.
Instead, we apply perturbation theory to both the ground and excited orbitals, resulting in an effective four-band model.
We discuss the details of this expansion in App.~\ref{app:four_band}.
For $|\tilde{\delta}_\epsilon|$ beyond the orbital inversions, we have calculated the excitation gap and give its results in Fig.~\ref{fig:energy_gap}.
We find that, in the regime $|\tilde{\delta}_\epsilon|\sim[5,10]$ ($R_c\sim 10 nm$ and $\sim 10\%$ lattice mismatch), the excitation gap is renormalized by up to $50\%$ with respect to the $[001]$ growth direction.
We conclude that, particularly in the low strain regime, a perturbative treatment of $H_c$ is insufficient.
Therefore, $H_c$ must be explicitly included in the perturbative basis.

\section{Magnetic response and effective g-tensor} \label{sec:g_fac}

\begin{figure}
    \centering
    \includegraphics[width=\linewidth]{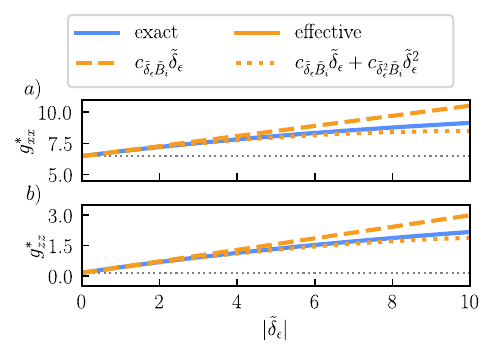}
    \caption{Exact diagonalization (black) and perturbation theory results (blue) for the parallel, $g^*_{zz}$, and perpendicular, $g^*_{xx}$, components of the $[001]$ $g^*$-tensor to different orders in perturbation theory. The strain exerted by the Bir-Pikus Hamiltonian, eq.\eqref{LK:BP}, changes the energy gap between the ground and excited orbitals. This yields a dependency of the $g^*$-tensor on the renormalized strain parameter, $\tilde{\delta}_\epsilon$.}
    \label{fig:gvsstrain}
\end{figure}

\begin{figure*}
    \centering
    \includegraphics[width=\linewidth]{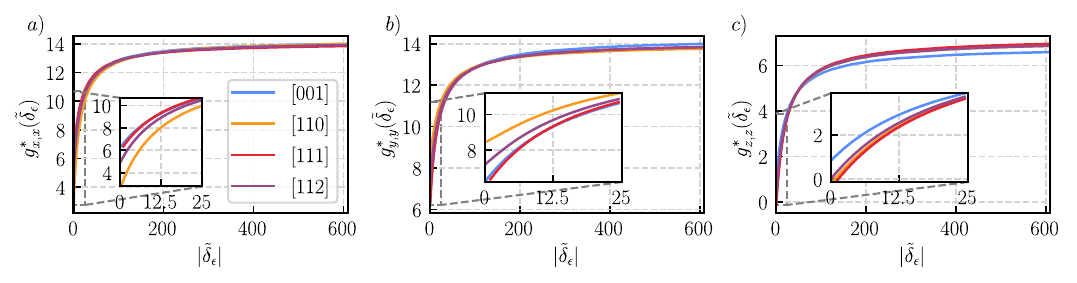}
    \caption{Effective g-factor, $g^*_{ii}(\tilde{\delta}_\epsilon)$ for crystal alignments along $[001], [110], [111], [112]$. We find that the g-factor for perpendicular fields is generically stronger that when applied in parallel to the wire, as expected from previous studies. The cubic term leads to an almost $50\%$ renormalization of the perpendicular g-factor for the $[110]$ direction compared to other directions. Similarly, the parallel g-factor of the $[001]$ direction obtains an appreciable finite value. All g-factor increase rapidly with the increasing excitation gap to approach their plateau values.}
    \label{fig:delta_gamma_of_strain}
\end{figure*}

\begin{figure}
    \centering
    \includegraphics[width=\linewidth]{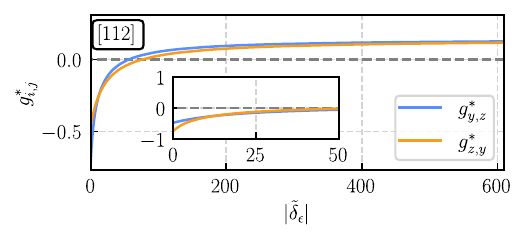}
    \caption{$g^*_{yz}$ and $g^*_{zy}$ components of the g-tensor for wires in the $[112]$ direction. We find these to become finite due to lack of rotational symmetry around the wire axis.}
    \label{fig:g_112_off_diag}
\end{figure}

Electrons and holes both couple to magnetic fields via the linear and cubic Zeeman terms as well as the orbital field.
Their lowest order response to magnetic fields is captured in the $g^*$-tensor.
Its associated energy scale, the Zeeman energy, is relevant in a wide variety of applications, e.g. resonance frequencies in spin qubits~\cite{nielsen_quantum_2010}, and topological superconductors~\cite{lutchyn_majorana_2010, laubscher_majorana_2024, laubscher_germanium-based_2024}.

In our approach, we identify the effective $g^*$-tensor with the coefficient of all terms linear in magnetic field strength in the perturbative expansion, i.e.
\begin{align}
    H_{Z, \rm  eff} = \frac{\mu_B}{2} g^*_{ij} B_i \sigma_j + \ldots.
\end{align}
 Its elements can be extracted through
\begin{align} \label{eq:g_tensor}
    g^*_{ij} = \operatorname{Tr}\left(\left.\frac{\partial H_{\rm eff}}{\partial \tilde{B}_i}\right|_{\tilde{B}_i=0} \sigma_j \right),
\end{align}
where the $g^*$-tensor implicitly depends on all other perturbations such as strain or electric fields.
In general, the entries of the $g^*$-tensor are restricted by the symmetries of the underlying system.
In the absence of strain, but with cubic terms included, the bulk system has full cubic $O_h$ symmetry.
This results in an isotropic bulk $g^*$-tensor.
A wire geometry breaks this symmetry group down to a subgroup depending on the wire orientation.
The symmetries of the growth directions we investigate significantly restrict the form of the $g^*$-tensor.
In our chosen basis (see Sec.~\ref{sec:model_pt}), the twofold rotation symmetry around $x$ with $R = \operatorname{diag}(1, -1, -1)$ ensures that $g^*_{xy}=g^*_{xz}=g^*_{yx}=g^*_{zx}=0$ for all studied crystal directions.
Rotation symmetry around the wire axis forces $g^*_{yz}=g^*_{zy}=0$ for $[001]$, $[011]$ and $[111]$ wires.
Moreover, the four and threefold rotation symmetries of $[001]$ and $[111]$ wires ensure $g^*_{xx} = g^*_{yy}$.
For details of this analysis, see Appendix~\ref{app_symmetry}.

\subsection{Perturbative treatment of strain}

Achieving quantitative agreement between full and effective models requires inclusion of sufficiently high perturbative orders.
We illustrate this requirement with strain as example by calculating higher orders in strain for wires in the $[001]$ growth direction.
We extract $g^*$ as the prefactor polynomial of the terms linear in the magnetic field components $B_i$.
We choose $H_0=H_{\odot}$ as the unperturbed Hamiltonian, neglect $H_c$, and use the basis states described in eq.~\eqref{eq:schroed_states}~\cite{kloeffel_strong_2011, kloeffel_direct_2018}.
Furthermore, we let $H^\prime=H_{BP}+H_Z+H_{LK, orb}$ (cf. Sec.~\ref{sec:model_pt}).
For the $[001]$ nanowire, the $g^*$-tensor is diagonal.
At higher orders, we find the polynomial expressions describing $g^*$ to be
\begin{align}\label{eq:gfactor_strain}
    g^*_{ii} &= 2(c_{\tilde{B}_i}+c_{\tilde{B}_i\tilde{\delta}_\epsilon}\tilde{\delta}_\epsilon+c_{\tilde{B}_i\tilde{\delta}_\epsilon^2}\tilde{\delta}_\epsilon^2),
\end{align}
where the factor $2$ is a consequence of our units.
Here, the lowest order coefficients, $c_{\tilde{B}_i}$, describe the bare g-factor in the absence of strain.
Our findings agree with previous results~\cite{kloeffel_strong_2011, adelsberger_enhanced_2022}.
In Fig.~\ref{fig:gvsstrain} we show the results for $g^*_{ii}$ up to second order in strain $\tilde{\delta}_\epsilon$, and compare the results of our perturbation theory against exact diagonalization of the full Hamiltonian.

To first order, the effective g-factor is overestimated.
This is due to the eigenstates of $H_\odot$.
The groundstate orbital has a significant weight on the $j_z=\pm 3/2$ states~\cite{kloeffel_strong_2011}.
This leads to a strong susceptibility to the $J_z^2$ term of $H_{BP}$.
To second order, the virtual transitions into the excited manifold correct this overestimation.
The strain correction is most important to include for magnetic fields parallel to the wire axis. 
As shown in Fig.~\ref{fig:gvsstrain}, the bare g-factor is nearly zero in this case. 
Therefore, any correction from strain is much larger than the bare result itself.
Our result reveals that a good representation of strain effects requires at least a second order perturbative expansion. 
This aspect will become relevant in Sec.~\ref{Sec:effective_mass}.

\subsection{Perturbative expansion at arbitrary strain} \label{subsec:strain_dep}

While the perturbative approach is valid for small strain or large level spacing, experimental systems often exhibit strong strain-induced renormalization.
As a result, the perturbative treatment of strain becomes invalid.
Growth studies have reported nanowires with shell radii of $\sim20\%$ for the core for $[111]$~\cite{nguyen_diameter-independent_2014} and $\sim40\%$ in $[110]$ and $[112]$~\cite{conesa-boj_boosting_2017, goldthorpe_inhibiting_2009}.
These correspond to $\tilde{\delta}_\epsilon \sim -0.14R_c^2/eV$ and $-0.21R_c^2/eV$, respectively.
Given the presented parameters~\cite{conesa-boj_boosting_2017, nguyen_diameter-independent_2014}, our theory suggests energy gaps $>20meV$ ($\tilde{\delta}_\epsilon\approx -14$ for $R_c=10nm$) between the lowest two orbitals.
As Fig.~\ref{fig:delta_gamma_of_strain} shows, lattice effects can lead to g-factor renormalizations of $20\%$ at these parameters.

We therefore incorporate the Bir-Pikus Hamiltonian directly into $H_0$ (see eq.~\eqref{eq:h_pert_symb}).
By fixing a reference point $|\tilde{\delta}_\epsilon|$, we obtain a well-defined low-energy basis for the perturbative expansion.
In this way, the generated effective coefficients depend on the $\tilde{\delta}_\epsilon$ reference point.
Deviations in strain away from the reference point can again be treated perturbatively. 
This allows us to extract expressions such as eq.~\eqref{eq:gfactor_strain}.
In Fig.~\ref{fig:strain_map} a) in App.~\ref{sec:strain_model}, we plot $\tilde{\delta}_\epsilon$ depending on core and shell radii for the core/shell ratios considered in this work.

In the remainder of this work we therefore choose $H_0=H_\odot+H_{BP}(\tilde{\delta}_\epsilon)+H_C$.
The $g^*$-tensor then becomes dependent on $\tilde{\delta}_\epsilon$ through the expansion point, and eq.~\eqref{eq:gfactor_strain} reduces to
\begin{align}
    g^*_{ij}(\tilde{\delta}_\epsilon)=2c_{\tilde{B}_i}(\tilde{\delta}_\epsilon).
\end{align}
As discussed in detail in App.~\ref{app_symmetry}, the $g^*$-tensor is diagonal in all growth directions but $[112]$.
Due to the lack of rotation symmetry around the wire axis, $[112]$ features finite $g^*_{yz}, g^*_{zy}$.
We show the results for $g^*_{ii}$ in Fig. \ref{fig:delta_gamma_of_strain} and the off-diagonal values $g^*_{yz}, g^*_{zy}$ for the $[112]$ direction in Fig. \ref{fig:g_112_off_diag}.
The results corroborate the generally known anisotropy between parallel and perpendicular g-factors~\cite{kloeffel_strong_2011,kloeffel_direct_2018, froning_single_2018}.
Our results show that this holds also for different growth directions of the nanowire.
Although the g-factor qualitatively retains its behavior, the renormalizations from the lattice misalignment are significant.
For the $[110]$ growth direction, we see a renormalization of $\sim50\%$ from the cubic symmetry terms.
For increasing $|\tilde{\delta}_\epsilon|$ the effective g-factor converges to limiting values of $\sim 14$ for perpendicular and $\sim 7$ for parallel magnetic fields.

\section{Linear and cubic spin-orbit interaction in the low-energy Hamiltonian} \label{sec:rashba}

In general, spin-orbit interaction (SOI) requires inversion symmetry breaking~\cite{winkler_spinorbit_2003}.
A well-known example is the Rashba SOI, which arises from structural inversion asymmetry in the crystal potential~\cite{winkler_spinorbit_2003}. 
In confined systems, additional SOI contributions can arise from externally induced symmetry breaking, for instance due to electric fields or heterostructure asymmetries~\cite{mauro_strain_2025}.
In Ge nanowires, one such mechanism is the direct Rashba spin–orbit interaction (DRSOI)~\cite{kloeffel_strong_2011}, which occurs when an external electric field breaks the cylindrical symmetry of the wire.

Similar to the $g^*$-tensor, the symmetries of the system determine which elements of the linear, $\doubleunderline{\alpha_{L}}$, and cubic SOI tensors, $\doubleunderline{\alpha_{C}}$, are finite.
This is determined by the relation between translational invariant axes, crystal symmetry, and applied fields. 
In our setting, we find that for the $[001]$ and $[111]$ growth directions, the only nonzero coefficient is $\alpha_{xy}=-\alpha_{yx}$.
In contrast, nanowires grown in $[110]$ can yield independent, finite $\alpha_{xy}$ and $\alpha_{yx}$.
Further, the $[112]$ orientation additionally yields finite $\alpha_{xz}$ and $\alpha_{zx}$ contributions.
We give a more detailed account of these observations in App. \ref{app_symmetry}.

In our perturbative analysis, SOI effects enter through the externally applied electric field~\cite{kloeffel_strong_2011}.
We are specifically interested in higher-order contributions in $k_z$ such as cubic-SOI.
We let $H_0=H_\odot+H_C+H_{BP}$ and $H^\prime=H_{k_z}+H_Z+H_{Z,3}+H_E$ and include the orbital field.
We then perform perturbation theory up to fourth order.
Contributions to the SOI tensors are given by
\begin{align} \label{eq:so_tensor}
    (\alpha_L)_{ij}&=\operatorname{Tr}\left(\frac{\partial^2 H_{eff}}{\partial k_z\partial\tilde{E}_i}\bigg|_{\tilde E_i=0} \sigma_j\right)\tilde{E}_i \\
    (\alpha_C)_{ij}&=\operatorname{Tr}\left(\frac{\partial^4 H_{eff}}{\partial k_z^3\partial\tilde{E}_i}\bigg|_{\tilde E_i=0} \sigma_j\right)\tilde{E}_i.
\end{align}
We find the most significant contributions to the linear and cubic SOI tensors to be
\begin{align}
    (\alpha_{L})_{ij} &= \left(c_{k_z\tilde{E}_i}(\tilde{\delta}_\epsilon)+\sum_lc_{k_z \tilde{E}_i \tilde{B}^2_l}(\tilde{\delta}_\epsilon) \tilde{B}_l \right)\tilde{E}_i, \label{eq:linear_rashba}\\
    (\alpha_{C})_{ij} &= c_{k_z^3\tilde{E}_i}(\tilde{\delta}_\epsilon)\tilde{E}_i, \label{eq:cubic_rashba}
\end{align}
\subsection{Linear spin-orbit interaction} \label{subsec:lin_rashba}

\begin{figure}
 \centering
 \includegraphics[width=\linewidth]{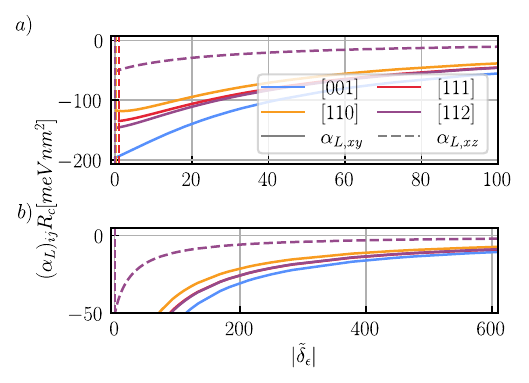}
 \caption{Linear Rashba coefficient for electric fields applied perpendicularly to the nanowire axis. The result highlights sensitivity of the effect to both, the energy gap, but also lattice alignment. For wires that are strained not far off an orbital inversion, the strength of the Rashba effect becomes sizable due to the enhanced coupling between states. It can also be seen that, due to the low symmetry of the $[112]$ grown crystal, the SOI has multiple components for that direction. In particular, a term parallel to the wire axis of about half the magnitude of the expected direction.}
 \label{fig:linear_rashba_x_combined}
\end{figure}

The bare linear SOI effect is mediated by virtual processes involving the excited orbital.
It is therefore inversely proportional to the excitation gap~\cite{kloeffel_direct_2018}.
Conversely, the maximally applicable electric field permitting a well-defined lowest orbital SOI is given by $|eR_cE_x|=E_{gap}$, i.e., the electric field associated with the excitation gap
This implies a competition of scales: the energy of the electric field scales $\sim R_c$, increasing spin-orbit.
The excitation gap, in turn, scales as $\sim R_c^{-2}$ and limits the maximally permissible field.
Engineering the strain of the nanowire, $E_{gap}\sim\tilde{\delta}_\epsilon\sim R_s/R_c$, therefore allows to tailor the range of electric field strengths.
Because this direct SOI effect is a consequence of the interplay between confinement and breaking of lattice symmetries, the question arises how the orientation of the crystal influences the magnitude of the spin-orbit effect.
We find that, perhaps surprisingly, the magnitude of the spin-orbit interaction decreases for crystal orientations other than $[001]$ by up to $30\%$ ($[110]$ orientation).
Naively we would expect larger SOI, as all directions other than $[001]$ feature an orbital inversion as described in Sec. \ref{sec:g_fac}, which reduces the excitation gap.
Our results, however, show that the electric field terms couple states less efficiently due to the competing symmetries in the system leading to the decrease.
Given results of material studies~\cite{conesa-boj_boosting_2017, nguyen_diameter-independent_2014, wang_epitaxial_2021} suggesting growth directions predominantly different from $[001]$, this is relevant.
In addition to the magnitude of the SOI, we also want to highlight the emergence of $\alpha_{L,xz}$ contributions of up to $\alpha_{L,xy}/\alpha_{L,xz}\sim 30\%$ for the $[112]$ crystal orientation. 
This is a consequence of the $[112]$ orientation lacking a second mirror plane that contains the wire axis which allows for more elements of the spin-orbit tensor, eq. \ref{eq:so_tensor} to be finite yielding the observed $\alpha_{L,xz}\sim E_x k_z$ contribution. 
In Fig.~\ref{fig:linear_rashba_x_combined} we show the maximally attainable spin-orbit strength, choosing $|eR_cE_x|=E_{gap}$.

The next correction to the linear SOI coefficient comes in quadratic order of the magnetic field. 
We generally find a negative (positive) trend for electric fields applied in the $x$ ($y$) direction, regardless of the direction of the magnetic field in the regime of small $\tilde{\delta}_\epsilon$. 
This is in agreement with Ref.~\cite{adelsberger_enhanced_2022}.
Increasing $|\tilde{\delta}_\epsilon|$, we however find that for the $[110], [111]$, and $[112]$ the sign of the correction term changes sign. 
We understand the behavior of the correction as a consequence of increasing gaps due to Zeeman coupling.
The opening of gaps between spins changes the magnitude of the virtual processes mediating the spin-orbit interaction for the two spin species.
Crucially, due to the induced gap, the different spin are affected differently, leading to the renormalization of the SOI.
Due to the dependence on the SOI processes itself, the renormalization process is most relevant in the limit of small $|\tilde{\delta}_\epsilon|$.

\subsection{Cubic spin-orbit interaction}

\begin{figure}
     \centering
     \includegraphics{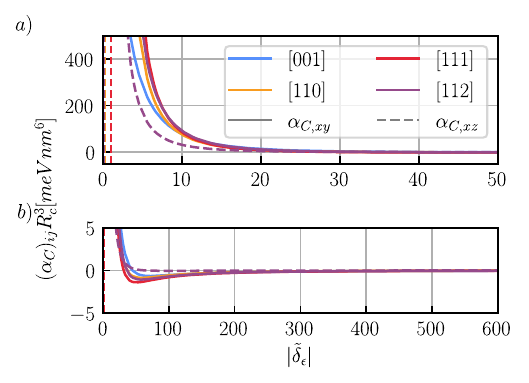}
     \caption{Effective cubic Rashba coefficient $\alpha_{R,C,x}$. Vertical dashed lines indicate locations of the orbital inversions for the lattice alignments. a) The cubic term takes take on appreciable values for a range of $\tilde{\delta}_\epsilon$. The implied larger separation of spins in parallel momentum could be of experimental use particularly in the low density limit. We again find a $\sigma_z$ contribution to the SOI in the $[112]$ direction of half the magnitude of the expected $\sigma_y$ term. b) Close-up of $\alpha_{R,C,x}$. We find that a cubic Rashba term features sign inversions for all lattice alignments.}
     \label{fig:cubic_rashba_x_combined}
 \end{figure}

We conclude the section with a discussion of the cubic SOI coefficient, eq.~\eqref{eq:cubic_rashba}, $\sim k_z^3$.
Cubic SOI terms come from an interplay between linear SOI and effective mass $\sim k_z^2/m^*$.
For each of the two processes, the excitation gap between the lowest two orbitals (cf.~\ref{fig:energy_gap}) must be limited.
This will be the case in proximity to the orbital inversions of the different lattice alignments, or negligible orbital spacing.
These situations should be experimentally feasible.
The results of~\cite{conesa-boj_boosting_2017} present wires featuring $R_c=5nm, R_s=7nm$, such that $|\tilde{\delta}_\epsilon|\approx 5.3$, which our analysis suggests is still within a regime where cubic SOI effects are relevant.
We present the results of our analysis in Fig.~\ref{fig:cubic_rashba_x_combined} for all studied growth directions.
We recognize that cubic SOI indeed becomes appreciable in regimes with limited excitation gaps.
 
\section{Effective mass, quasi flat bands, and their spin dependency} \label{Sec:effective_mass}

\begin{figure*}
    \centering
    \includegraphics[width=\linewidth]{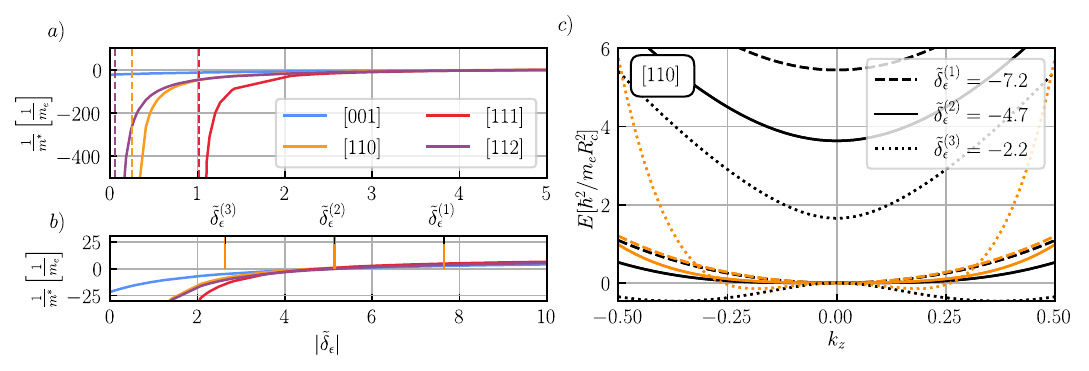}
    \caption{Inverse effective mass in dependence of strain and sub-band energies in dependence of $k_z$. a) Inverse effective mass $1/m^*$ as function of strain. The plot reveals the relevance of the energy gap for the convergence of the inverse effective mass in perturbation theory. Vertical dashed lines correspond to previously determined strains of orbital inversions between the lowest two sub-band orbitals. b) Close-up of the inverse effective mass. We recognize the presence of a divergence indicated at $\tilde{\delta}_\epsilon^{(2)}$. The exact location of the divergence only differs slightly between the different lattice alignments. c) Energy spectrum from exact diagonalization (black) and the effective Hamiltonians (orange) for the $[110]$ growth direction depending on $k_z$ for three values of $\tilde{\delta}_\epsilon$ in the vicinity of the divergence. Indeed, the energy bands show a relatively flat band over a broader range of $k_z$. In situations of low hole-filling, these points of divergence might yield properties useful for correlated states.}
    \label{fig:effective_mass}
\end{figure*}

Finally, we turn our attention to the effective mass.
Previous literature suggests that the effective mass is strain dependent~\cite{kloeffel_direct_2018, adelsberger_enhanced_2022}.
Further, it has been shown that the effective mass couples to the magnetic field in high-orders~\cite{adelsberger_enhanced_2022}.
This raises the question of in-situ tuning of the effective mass by externally applied fields.
We answer this question by studying the effective mass
\begin{align} \label{eq:eff_mass_trace}
    \frac{1}{m^*}=\Tr\left[\frac{\partial^2 H_{eff}}{\partial k_z^2}\bigg|_{k_z=0}\sigma_0\right],
\end{align}
and its spin-dependent correction,
\begin{align} \label{eq:eff_mass_cor_trace}
    \frac{\partial}{\partial B_i}\left[\frac{1}{m^*}\right] = \Tr\left[\frac{\partial^3 H_{eff}}{\partial k_z^2 \partial \tilde{B}_i}\bigg|_{k_z=0, B_i=0} \sigma_i\right].
\end{align}
We find their most significant contributions to be captured by the polynomials
\begin{align}
\begin{split}
\frac{1}{m^*} &= c_{\tilde{k}_z^2}(\tilde{\delta}_\epsilon) \\
&+\sum_i c_{\tilde{k}_z\tilde{B}_i^2}(\tilde{\delta}_\epsilon)\tilde{B}_i^2 
  + c_{\tilde{k}_z\tilde{E}_i^2}(\tilde{\delta}_\epsilon)\tilde{E}_i^2
\label{eq:eff_mass}
\end{split}
\\
\begin{split}
\frac{\partial}{\partial B_i}\!\left[\frac{1}{m^*}\right] 
&= c_{\tilde{k}_z^2\tilde{B_i}}(\tilde{\delta}_\epsilon)  \\
&+c_{\tilde{k}_z^2\tilde{B}_i\tilde{\delta}_\epsilon^\Delta}(\tilde{\delta}_\epsilon)\tilde{\delta}_{\epsilon}^\Delta
  +c_{\tilde{k}_z^2\tilde{B}_i(\tilde{\delta}_\epsilon^{\Delta})^2}(\tilde{\delta}_\epsilon)(\tilde{\delta}_{\epsilon}^\Delta)^2
\label{eq:eff_mass_cor}
\end{split}
\end{align}
where $\tilde{\delta}_\epsilon^\Delta$ is the deviation in strain away from the reference point $\tilde{\delta}_\epsilon$.

\subsection{Divergent effective mass and quartic bands}

Our first observation is that the bare effective mass, eq.~\eqref{eq:eff_mass} changes its sign depending on strain.
This is visible for each growth direction in Fig.~\ref{fig:effective_mass} 
This implies the existence of a point for which the effective mass diverges leading to quasi flat orbitals and higher order van-Hove singularities with quartic dependence.
This can potentially be interesting for the study of correlated states in such strained nanowires systems.
In Fig.~\ref{fig:effective_mass} a) and b), we show the $|\tilde{\delta}_\epsilon|$ dependence of the bare effective mass for all crystal directions.
Interestingly, the point where the effective mass diverges does not differ significantly between the different lattice alignments.
To corroborate our finding, we plot the energy bands calculated by exact diagonalization for three values of $\tilde{\delta}_\epsilon$ around the divergence for a wire grown in $[110]$ in Fig.~\ref{fig:effective_mass} c).
Indeed we find that, at the point of the effective mass divergence, the energy band is flat over a broad range of momenta.

Furthermore, we find that electric and magnetic fields allow to tune the effective mass in-situ.
We discuss the results of these calculations in App.~\ref{app:eff_mass}.

\subsection{Magnetically tuned quartic bands}

\begin{figure}
    \centering
    \includegraphics[width=\linewidth]{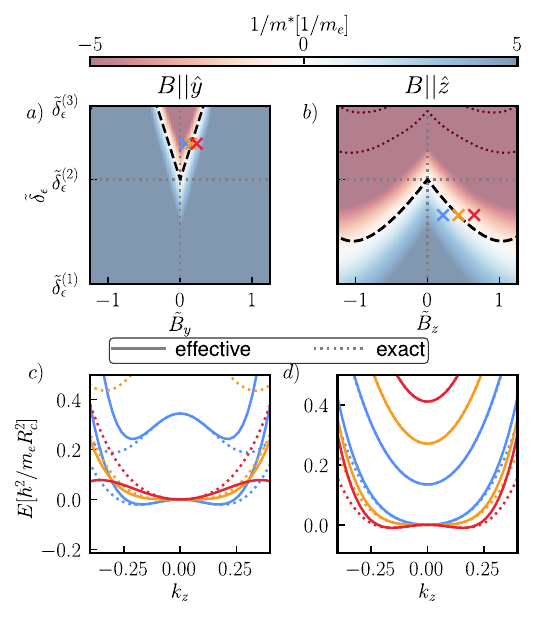}
    \caption{Vicinity of the strain where the effective mass diverges against magnetic fields in both, $\hat{y}$ and $\hat{z}$ direction. a) and b) demonstrate the existence of points given by the solutions of eq.~\eqref{eq:eff_mass_parametrization} (black dashed lines) where the effective mass diverges by application of external magnetic fields. Given the positive sign of the g-factors, this means that one obtains a single bands with quartic dispersion with positive (negative) effective spin given positive (negative) magnetic fields. c) and d) show the bands calculated both, using the effective model, and exact diagonalization corroborating our analysis.}
    \label{fig:effective_mass_tuning}
\end{figure}

Beyond the renormalization, the effective mass also couples to the magnetic field. 
This proportionality yields two distinct effects.
First, if the strain where the mass diverges is targeted, but not attained in growth, magnetic and electric fields allow to renormalize the effective mass to the desired divergent value.
Second, tuning the magnetic field alone allows to tune to points where a single, spinful level constitutes the ground state and features a divergent effective mass.
This ground state is then separated by a gap proportional to both the distance $\tilde{\delta}_\epsilon$ from the divergent point and the splitting gap $\partial_{B_i}[1/m^*]|_{B_i=0}B_i$.
Our effective models allow to extract the parametrization of these lines which take the form
\begin{equation} \label{eq:eff_mass_parametrization}
\begin{split}
    0&=\frac{1}{m^*}\bigg|_{\tilde{E}_i=0} + \frac{\partial}{\partial B_i}\left[\frac{1}{m^*}\right] \\
    &=\sum_{i,j} c_{\tilde{k_z}^2 (\tilde{\delta}_{\epsilon}^\Delta)^i \tilde{B}_d^j}(\tilde{\delta}_\epsilon^\Delta)^i \tilde{B}_d^j,
\end{split}
\end{equation}
where $d=x,y,z$ corresponds to the orientation of the magnetic field.
We want to stress that, in connection with the positive sign of the effective g-factor, the sign of the magnetic field determines the effective spin polarization of the tuned, quartic band.
To illustrate our discussion we plot our results for the $[110]$ direction in Fig.~\ref{fig:effective_mass_tuning}.
To allow to compare, we focus on the same range of $\tilde{\delta}_\epsilon$ as in Fig.~\ref{fig:effective_mass}.
Similarly, we take three exemplary points for which we compare the energy bands of the effective model with exact diagonalization.
Not only do we find good agreement between our effective model and exact diagonalization but also we find that our effective theory indeed well reproduces our assertion. 

\section{Conclusion} \label{sec:conclusion}

We have studied effective single orbital models of Ge/Si core/shell nanowires in a perturbation theory approach.
Our approach allows to study effective coefficients such as g-factors and spin-orbit interactions, but also their renormalizations through external fields by expanding to higher orders.
We explored the role and effect of the lattice orientation-dependent cubic terms on the electronic structure.
In the limit of low strain or small level spacing, the cubic terms introduce orbital inversions which make them relevant to include in perturbative expansions.
An analysis of the effective g-factors for the different growth directions shows renormalization effects lead to a $\sim50\%$ change of the g-factors when compared to the $[001]$ direction.
The perturbative approach allowed us to resolve different influences of both linear and cubic spin-orbit terms.
We showed sizable renormalizations due to different growth directions and the perhaps counterintuitive consequences this has with respect to linear spin-orbit effects.
We demonstrated how in systems with small excitation gaps, the cubic spin-orbit interactions become significant.
Moving to the effective mass, we showed the existence of points which feature divergent effective mass, making the lowest occurring order of momentum being quartic in the absence of fields.
We expect these points to be interesting for physics of correlations due to their high density.
Lastly, we showed how this property can be used to tune to single polarized quartic bands in dependence of magnetic fields and strain.
Given current progress on proximity superconductivity in Ge~\cite{pino_theory_2025, babkin_superconducting_2025} and recent experimental implementations~\cite{wu_magnetic_2024, ridderbos_hard_2020, ridderbos_multiple_2019}, we expect our results to be useful for the pursuit of Ge systems hosting e.g., Majorana fermions~\cite{laubscher_majorana_2024, laubscher_germanium-based_2024}, or tunable Andreev spin qubits~\cite{chtchelkatchev_andreev_2003}.

\section{Acknowledgements}

The authors acknowledge useful discussions with A.R. Akhmerov.
S.M. and A.M.B. acknowledge funding through the Dutch Organization for Scientific Research (NWO) through OCENW.GROOT.2019.004.
D.~V. was supported by the Deutsche Forschungsgemeinschaft (DFG, German Research Foundation) under Germany’s Excellence Strategy through the W\"{u}rzburg-Dresden Cluster of Excellence on Complexity and Topology in Quantum Matter – \textit{ct.qmat} (EXC 2147, project-id 392019),
the National Research, Development and Innovation Office of Hungary under OTKA grant no. FK 146499,
the János Bolyai Research Scholarship of the Hungarian Academy of Sciences,
and by the European Union within the Horizon Europe research and innovation programme via the ONCHIPS project under grant agreement No 101080022.

\emph{Author contributions}
D.V and M.W. conceived the project. S.M. and M.W. defined the scope of the project. S.M., A.M.B., and D.V. performed the calculations and interpretation under supervision of M.W.. S.M. and D.V. wrote the manuscript with input from all authors.

\emph{Data availability}
The code written to generate the data and all plots can be found in the Zenodo repository of Ref.~\cite{zenodo}.

\bibliography{references}

\appendix

\section{Hilbert-Schmidt transformation} \label{app:operator_transform}

Following the ideas in Refs.~\cite{maurer_multilevel_2020} and ~\cite{kimura_bloch_2003} we represent our effective Hamiltonians in a basis of operators, $\{Id\}\cap \{\hat{b}_i\}_{i=1,..N^2-1}$, where $N$ is the dimension of the Hilbert space.
We choose the $b_i$ to satisfy

\begin{align} \label{eq:op_norm}
    Tr[\hat{b}_i \hat{b}_j] &= \delta_{i,j} \\
    Tr[\hat{b}_i] &= 0.
\end{align}

A suitable choice of operators for our problem are the Pauli matrices and Kronecker-products thereof for multi-orbit effective models (cf. Sec. \ref{app:four_band})~\cite{kimura_bloch_2003, maurer_multilevel_2020},
We expand the effective Hamiltonian in the chosen basis through Hilbert-Schmidt inner products as

\begin{align} \label{eq:hs_decomp}
    H=\frac{\Tr[H]}{N}Id+\frac{1}{N}\sum_{i=1}^{N^2-1}\Tr[H \hat{b}_i] \hat{b}_i,
\end{align}

where the factor $1/N$ is a consequence of the normalization introduced in eq.~\eqref{eq:op_norm}.
For Hamiltonians depending on a multiple parameters $\{\lambda_i\}$, i.e. $H=H(\vec{\lambda})$, and we expand in each, an effective model will generally take the form

\begin{align} \label{eq:gen_eff_model}
    H_{\mathrm{eff}} = \sum_{i} p_i(\vec{\lambda})h_{i},
\end{align}

where $p_i(\vec{\lambda})$ are polynomials depending on the expansion parameters and the index $i$ is restricted to only allow polynomials until a maximally defined expansion order in each $\lambda_i$.
Performing a Hilbert-Schmidt transformation on the effective Hamiltonian yields the form

\begin{align} \label{eq:gen_hilbert_schmidt_model}
    H_{\mathrm{eff}} = \sum_{i,n}p_i(\vec{\lambda})c_{p_i(\vec{\lambda})}^{(n)}\frac{1}{N} \hat{b}_n.
    \end{align}

This Hilbert-Schmidt form has the advantage that, by choosing a useful set of basis operators such as the Pauli matrices, the effective Hamiltonians become very simple to interpret. 
As outlined in Sec.\ref{sec:model_pt}, a single, spinful orbital behaves effectively like a spin-1/2, despite its light- and heavy-hole mixing.
Representing the effective Hamiltonian then in the form given in eq.~\eqref{eq:gen_hilbert_schmidt_model} allows to understand the interesting regimes it contains by only analyzing the coefficients $c_{p_i{\vec{\lambda}}}^{(n)}$.
The code repository associated with this work, Ref.~\cite{zenodo}, in a first step, generates effective Hamiltonians of the form given in eq.~\eqref{eq:gen_eff_model}.
In a second step, we perform the transformation to the Hilbert-Schmidt form, eq.~\eqref{eq:gen_hilbert_schmidt_model}.

\section{Symmetry analysis of effective coefficients} \label{app_symmetry}

\subsection{General considerations}

The symmetries of the nanowire pose constraints on the coefficients in the effective Hamiltonian describing external perturbations.
Taking the active view of symmetry transformations~\cite{varjas_qsymm_2018}, we consider the action of a symmetry operator from the symmetry group of the system $g \in G$, and demand that the transformed system behave identically to the original one.
Without external perturbations, the symmetry constraint reads:
\begin{equation}
    g H = H,
\end{equation}
which in $\vb{k}$-space translates to
\begin{equation}
    (g H)(\vb{k}) = U H(R^T \vb{k}) U^{\dag} = H(\vb{k}),
\end{equation}
where $R$ is the real space orthogonal rotation action of $g$ and $U$ is its unitary action on the internal (spin or orbital) degrees of freedom.

Adding a dependence on external electromagnetic fields that transform as (pseudo) vectors, changes the symmetry constraint to
\begin{equation}
    g[H(\vb{E}, \vb{B})] = H(R \vb{E}, \overline{R} \vb{B}),
\end{equation}
where $\overline{R} = (\det R) R$ is the rotation action on axial vectors.
This relation expresses the fact that acting with a symmetry of the unperturbed system in the presence of symmetry-breaking perturbations results in the same system in the presence of the rotated perturbations.
In $\vb{k}$-space this is summarized as
\begin{equation}
    \label{eqn:symm}
    U H(R^T \vb{k}, R^T \vb{E}, \overline{R}^T \vb{B}) U^{\dag} = H(\vb{k}, \vb{E}, \vb{B}).
\end{equation}
In the 2-band effective models that we consider, we chose the basis of the low-energy subspace, such that the vector of $\sigma$ matrices transforms as an axial vector:
\begin{equation}
    U^{\dag} \sigma_i U = \overline{R}_{ij} \sigma_j.
\end{equation}

In the next two sections, we provide details for deriving the symmetry constraints on the effective $g$-tensor and the Rashba coefficients for the wire orientations studied in the paper.
The results are summarized in Table~\ref{tab:symmetry}.

\begin{table}[h]
\centering
\begin{tabular}{|c|Sc|Sc|}
\hline
\textbf{Wire orientation} & $g^*$ & $\alpha$ \\ \hline\hline
$\begin{matrix}
[001] \\
\phantom{1} \\
[111] 
\end{matrix}$ &
$\begin{pmatrix} g^*_{xx} & 0 & 0 \\
                 0 & g^*_{xx} & 0 \\
                 0 & 0 & g^*_{zz}
    \end{pmatrix}$ &
$\begin{pmatrix} 0 & \alpha_{xy} &  0 \\
                 -\alpha_{xy} & 0 & 0 \\
                 0 & 0 & 0
    \end{pmatrix}$ \\ \hline

$[110]$ &
$\begin{pmatrix} g^*_{xx} & 0 & 0 \\
                 0 & g^*_{yy} & 0 \\
                 0 & 0 & g^*_{zz}
    \end{pmatrix}$ &
$\begin{pmatrix} 0 & \alpha_{xy} &  0 \\
                 \alpha_{yx} & 0 & 0 \\
                 0 & 0 & 0
    \end{pmatrix}$\\ \hline
$[112]$ &
$\begin{pmatrix} g^*_{xx} & 0 & 0 \\
                 0 & g^*_{yy} & g^*_{yz} \\
                 0 & g^*_{zy} & g^*_{zz}
    \end{pmatrix}$ &
$\begin{pmatrix} 0 & \alpha_{xy} &  \alpha_{xz} \\
                 \alpha_{yx} & 0 & 0 \\
                 \alpha_{zx} & 0 & 0
\end{pmatrix}$\\ \hline
\end{tabular}
\caption{Symmetry constrained $g^*$ and $\alpha$ tensors for the various wire orientations consider.}
\label{tab:symmetry}
\end{table}

\subsection{Symmetry constraints on the $g$-factor}
Substituting the symmetry relation \eqref{eqn:symm} into the formula \eqref{eq:g_tensor} for the effective $g$-tensor
\begin{align} 
    g^*_{ij} = \operatorname{Tr}\left(\left.\frac{\partial H_{\rm eff}}{\partial \tilde{B}_i}\right|_{\tilde{B}_i=0, \vb{k}=0} \sigma_j \right),
\end{align}
and using the transformation properties of the $\sigma$ matrices we find
\begin{equation}
    \overline{R}^T g^* \overline{R} = g^*.
\end{equation}
Because the same sign appears twice for orientation-reversing symmetries, this is equivalent to $R^T g^* R = g^*$.

In the following, we use a coordinate system where the $z$ axis is aligned with the wire axis, and the $xy$ axes are perpendicular to it.
We use the $[hkl]$ notation to specify crystallographic directions relative to the cubic axes of the crystal. 
A twofold rotation around an axis perpendicular to the wire axis, inversion symmetry, and the mirror given by their combination are symmetries present in all of the wire directions we consider.
Aligning the $x$ axis with this rotation axis, the corresponding rotation matrix is $C_{2x} = -\overline{C_{2x}} = \rm{diag}(1, -1, -1)$.
This enforces $g^*_{xy}=g^*_{xz}=g^*_{yx}=g^*_{zx}=0$ in all cases.
Inversion symmetry does not result in any constraints on the $g$-tensor.
In the $[112]$ oriented wire the $x$ axis is in the $[1\overline{1}0]$ direction, and it does not have any further symmetries.

The $[011]$-oriented wire has two additional twofold rotation symmetries besides the $x=[100]$ axis, around the $y=[0\overline{1}1]$ and $z=[011]$ axes, given by $C_{2y} = -\overline{C_{2y}} = \rm{diag}(-1, 1, -1)$ and $C_{2z} = \overline{C_{2z}} = \rm{diag}(-1, -1, 1)$.
One of these is sufficient to ensure $g^*_{yz}=g^*_{zy}=0$, hence a diagonal effective $g$-tensor.

In the $[111]$ oriented wire $x=[10\overline{1}]$, and it also has a threefold rotation axis around the wire axis given by
\begin{equation}
    C_{3z} = \overline{C_{3z}}= \begin{pmatrix} \phantom{-}c & s & 0 \\
                            -s & c & 0 \\
                            \phantom{-}0 & 0 & 1
    \end{pmatrix},
\end{equation}
where $c = \cos 2\pi/3$ and $s = \sin 2\pi/3$.
This enforces $g^*_{yz}=g^*_{zy}=0$ and $g^*_{xx} = g^*_{yy}$.

Finally, the $[001]$ wire has a fourfold rotation symmetry around its axis given by
\begin{equation}
    C_{4z} = \overline{C_{4z}} = \begin{pmatrix} \phantom{-}0 & 1 & 0 \\
                            -1 & 0 & 0 \\
                            \phantom{-}0 & 0 & 1
    \end{pmatrix}.
\end{equation}
Similarly to the previous case, this enforces $g^*_{yz}=g^*_{zy}=0$ and $g^*_{xx} = g^*_{yy}$.

\subsection{Symmetry constraints on the Rashba coefficients}
Let us write the most general Rashba-like term as
\begin{equation}
    H_R = \alpha_{ij} k_z E_i \sigma_j.
\end{equation}
Substituting into the symmetry condition \eqref{eqn:symm} we find
\begin{equation}
    R_{zz} R^T \alpha \overline{R} = \alpha.
\end{equation}
Here we used the fact that any symmetry of the wire has to leave the wire axis invariant, hence $R_{zz} = \pm 1$.
The same constraints apply to the cubic Rashba terms, as $k_z^3$ transforms the same way as $k_z$.

The $C_{2x}$ symmetry present in all wire directions we consider forces $\alpha_{ii} = 0$ and $\alpha_{yz} = \alpha_{zy} = 0$.
As the $[112]$ wire does not have other constraints, this allows for terms of the form $k_z E_z \sigma_x$ and $k_z E_x \sigma_z$, furthermore, the terms $k_z E_y \sigma_x$ and $k_z E_x \sigma_y$ do not necessarily appear with opposite coefficients like they would in $k_z (\vb{E}\times \sigma)_z$.

The $C_{2y}$ symmetry of the $[011]$-oriented wire further enforces $\alpha_{xz} = \alpha_{zx}=0$, however, $\alpha_{xy} \neq -\alpha_{yx}$ still in this case.

The $C_{3z}$ or $C_{4z}$ symmetries present in $[111]$ and $[001]$ wires, respectively, are sufficient to ensure $\alpha_{xy} = -\alpha_{yx}$ besides the constraints above, resulting in the standard form of the Rashba term $k_z (\vb{E}\times \sigma)_z$.

\section{Effective models for two orbitals} \label{app:four_band}

\begin{figure*}
    \centering
    \includegraphics[width=\linewidth]{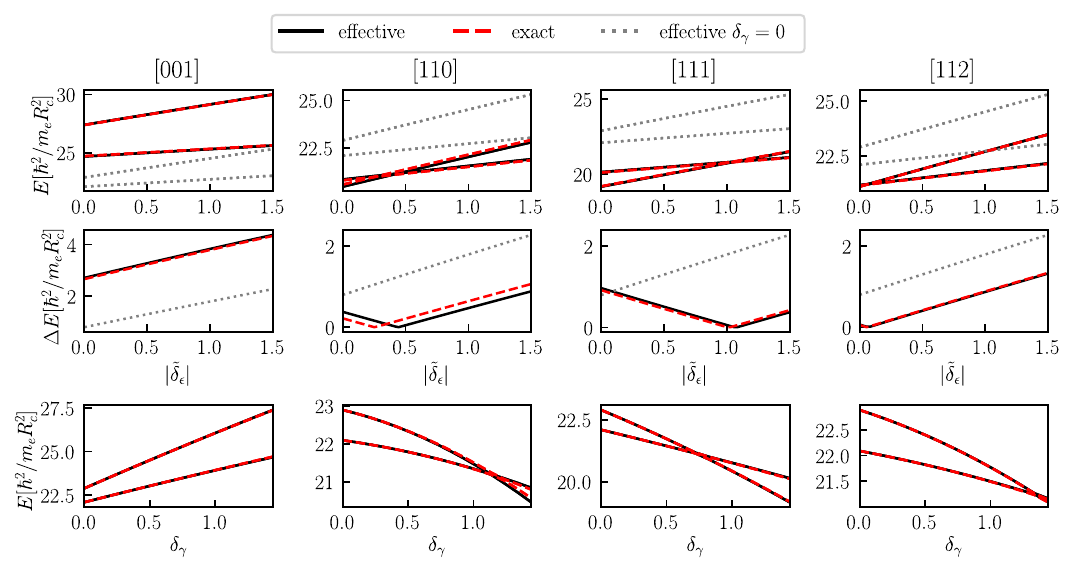}
    \caption{Energy vs. strain parameter for the lowest two orbitals. Black solid line correspond to exact diagonalization while red dashed lines correspond to the two-orbital effective Hamiltonian. We recognize that a two-orbital expansion can still accurately represent the effects of the orbital inversion, making it a good description of the low energy subspace in the vicinity of them. With the gray dotted lines, we indicate the eigenvalues of the effective Hamiltonian based on an expansion taking only the spherical approximation, eq.~\eqref{LK:odot}, into account for the basis and neglecting eq.~\eqref{eq:cubic_term}. Note that, in the top row, $\delta_\gamma$ is taken into account perturbatively leading to the apparent discrepancy in the $[110]$ growth direction.}
    \label{fig:two_orbital_sweep}
\end{figure*}

In Sec. \ref{sec:g_fac} we elaborated on the occurrence of orbital inversions due to the terms associated with cubic symmetries.
In the vicinity of these points, obtaining a single orbital effective model is impossible due to the lack of an energy gap between ground and excited orbital.
Instead, one can however perform a perturbative expansion for both orbitals.
Doing so yields models with four levels described by two sets of Pauli matrices (see Sec.\ref{app:operator_transform}), $\tau, \sigma$, where $\tau$ corresponds to the orbital degree of freedom, and $\sigma$ to the effective spin degree of freedom~\cite{kloeffel_strong_2011, kloeffel_direct_2018}. 
For each crystal direction, we choose $H_0=H_{\odot}$ and treat all remaining terms in perturbation theory.
We restrict our discussion here to field alignments $E_{\perp}=\tilde{E}_x$ and $B_\perp = \tilde{B}_y, B_{||} = \tilde{B}_z$.
We furthermore suppress global energy offsets.
The resulting effective model consists of two distinct parts: a spherically invariant part and a part that depends on the alignment of the crystal with the wire proportional to $\delta_\gamma = \gamma_3-\gamma_2$ (cf. Sec.~\ref{sec:model_pt}).
The spherically invariant part, we restrict the expansion to second order in $k_z,\tilde{\delta}_\epsilon, \tilde{E}_i. \tilde{B}_i$.
We obtain
\begin{equation}
\begin{split}
        H_{\mathrm{eff}}^{(4,\odot)} &= \left(6.82 \tau_0\sigma_0 + 1.48 \tau_3\sigma_0\right)k_z^2 \\
    &-7.23\tau_2\sigma_1 k_z\\
    &-0.15\tau_1\sigma_3 E_\perp\\
    &+ \left(0.40\tau_0\sigma_3-0.44\tau_3\sigma_3\right)B_{||}\\
    &- \left(0.22\tau_0\sigma_2+1.40\tau_3\sigma_2\right)B_\perp\\
    &-2.65\tau_1\sigma_3 B_\perp k_z\\
    &-1.70\tau_1\sigma_2 B_{||}k_z\\
    &+ \left(-1.14\tau_0\sigma_0+0.51\tau_3\sigma_0\right)\tilde{\delta}_\epsilon\\
    &+ \left(-0.16\tau_2\sigma_1\right)\tilde{\delta}_\epsilon k_z, 
\end{split}
\end{equation}
and we find good agreement to previous results~\cite{kloeffel_strong_2011}.
Induced by the cubic symmetry terms, we find a $\delta_\gamma$ dependent part where the coefficients change with the crystal orientation.
We find
\begin{equation} \label{eq:four_band_cubic}
\begin{split}
        H_{\mathrm{eff}}^{(4,C)}&=\left(c_{\delta_\gamma;0,0} + c_{\delta_\gamma;3,0}  \right)\delta_\gamma  \\
            &+\left(c_{\delta_\gamma k_z, 2,1} + c_{\delta_\gamma k_z;1,0}  \right)\delta_\gamma k_z  \\
            &+\left(c_{\delta_\gamma k_z^2; 0,0} + c_{\delta_\gamma k_z^2; 3,0}  \right)\delta_\gamma k_z^2  \\
            &+\left(c_{\delta_\gamma B_{||}; 0,||} + c_{\delta_\gamma B_{||};3,||}  \right)\delta_\gamma B_{||}  \\
            &+\left(c_{\delta_\gamma B_{||} k_z; 1,\perp} + c_{\delta_\gamma B_{||} k_z; 1,||}   \right)\delta_\gamma B_{||} k_z  \\
            &+\left(c_{\delta_\gamma B_\perp; 0,\perp} + c_{\delta_\gamma B_\perp;3,\perp}    \right)\delta_\gamma B_\perp  \\
            &+\left(c_{\delta_\gamma B_\perp k_z; 1,\perp} +  c_{\delta_\gamma B_\perp k_z; 1,||}  \right)\delta_\gamma B_\perp k_z, 
\end{split}
\end{equation}
where the explicit coefficients can be found in Tab.~\ref{tab:four_band_coeffs}.

\begin{table}[]
    \centering
    \begin{tabular}{c|c|c|c|c}
         & $[001]$ & $[110]$ & $[111]$ & $[112]$   \\ \hline
       $c_{\delta_\gamma;0,0}$                & $2.60$  & $-0.65$ & $-1.74$ & $-0.65$   \\ \hline
       $c_{\delta_\gamma;3,0}$                & $-0.69$ & $0.17$ & $0.46$ & $0.17$   \\ \hline
       $c_{\delta_\gamma k_z, 2,1}$           & $-1.04$ & $0.26$ & $0.69$ & $0.26$   \\ \hline
       $c_{\delta_\gamma k_z;1,0}$                & $0$     & $0$    & $0$    & $1.44$  \\ \hline
       $c_{\delta_\gamma k_z^2; 0,0}$         & $-0.97$ & $0.24$ & $0.64$ & $0.24$   \\ \hline
       $c_{\delta_\gamma k_z^2; 3,0}$         & $0.10$ & $-0.02$ & $-0.06$ & $-0.02$   \\ \hline
       $c_{\delta_\gamma B_{||}; 0,||}$       & $0.02$ & $0.00$ & $-0.01$ & $-0.01$   \\ \hline
       $c_{\delta_\gamma B_{||};3,||}$        & $-0.17$ & $0.04$ & $0.11$ & $0.04$   \\ \hline
       $c_{\delta_\gamma B_{||} k_z; 1,\perp}$    & $-0.13$ & $0.03$ & $0.09$ & $0.03$   \\ \hline
       $c_{\delta_\gamma B_{||} k_z; 1,||}$    & $0$     & $0$    & $0$    & $0.03$  \\ \hline
       $c_{\delta_\gamma B_\perp; 0,\perp}$   & $0.15$ & $-0.75$ & $-0.01$ & $-0.28$   \\ \hline
       $c_{\delta_\gamma B_\perp;3,\perp}$    & $-0.18$ & $0.20$ & $0.12$ & $0.01$   \\ \hline
       $c_{\delta_\gamma B_\perp; 0,||}$    & $0$     & $0$    & $0$    & $-0.14$  \\ \hline
       $c_{\delta_\gamma B_\perp; 3,||}$    & $0$     & $0$    & $0$    & $0.36$  \\ \hline
       $c_{\delta_\gamma B_\perp k_z; 1,\perp}$   & $0.32$ & $-2.66$ & $-0.22$ & $0.26$   \\ \hline
       $c_{\delta_\gamma B_\perp k_z; 1,||}$   & $0$ & $0$ & $0$ & $-0.3$   \\ \hline
    \end{tabular}
    \caption{Relevant coefficients for the cubic part of the effective two-orbital Hamiltonians, eq.~\eqref{eq:four_band_cubic}.}
    \label{tab:four_band_coeffs}
\end{table}

Our results highlight how the cubic anisotropies lead to changes in all relevant parameters.
Most strikingly, it leads to an orbital inversion that gets lifted only for finite strain.
To illustrate this statement, we plotted the energies from exact diagonalization and diagonalizing the effective model in dependence of $\delta_\gamma$ in Fig.~\ref{fig:two_orbital_sweep}.
We also see how the alignment with the wire renormalizes the effective $g^*$-factor of the system which explains why the results presented in Fig.~\ref{fig:delta_gamma_of_strain} differ from e.g.~\cite{kloeffel_strong_2011}.
Finally, Fig.~\ref{fig:two_orbital_sweep} also illustrates why a perturbative of the cubic terms only can yield problems in subsequent calculations due to the incurred differences of the basis (cf. $[110]$).

\section{Strain model} \label{sec:strain_model}

\begin{figure*}
    \centering
    \includegraphics[width=\linewidth]{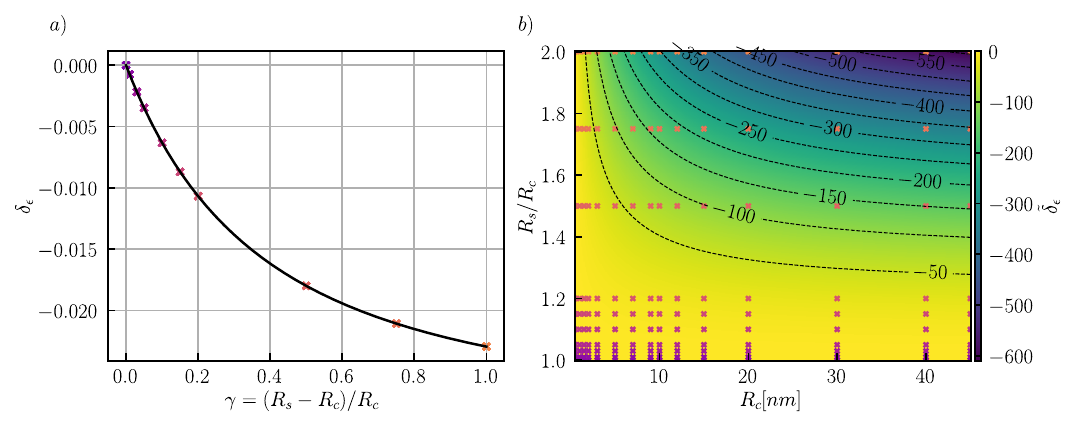}
    \caption{Bare ($\delta_\epsilon$ and rescaled strain ($\tilde{\delta}_\epsilon$) depending on core and shell radii. a) shows the bare strain and its dependence on core and shell radii via the parameter $\gamma$, detailed in~\cite{kloeffel_acoustic_2014}. Crosses show the core/shell ratios sampled for the effective models. b) shows the rescaled strain for the range of analyzed core radii and core/shell ratios. Crosses indicate the points sampled for the generation of the perturbative models.}
    \label{fig:strain_map}
\end{figure*}

The strain model used in this work is that developed in~\cite{kloeffel_acoustic_2014}.
A critical parameter for our analysis is $\delta_\epsilon=\epsilon_{zz}-\epsilon_\perp$, where $\epsilon_\perp=\epsilon_{xx}=\epsilon_{yy}$, and its rescaled partner, $\tilde{\delta}_\epsilon = \delta_\epsilon/E_0$.
In this section, we show the dependency of the bare ($\delta_\epsilon$) and rescaled strain $\tilde{\delta}_\epsilon$ on the core and shell radii.
In Fig.~\ref{fig:strain_map} we show the bare strain depending on the parameter $\gamma=(R_s-R_c)/R_c$ introduced in~\cite{kloeffel_acoustic_2014}.
The strain induced by the lattice constant mismatch is approximated to be applied homogeneously to the cylindrical core.
Fig.~\ref{fig:strain_map} also shows the rescaled strain, $\tilde{\delta}_\epsilon$ we use throughout the manuscript. 
Crosses in Fig.~\ref{fig:strain_map} b) indicate all sample points for which we generate effective models via our two level perturbation theory approach.

\section{Additional results for the effective mass} \label{app:eff_mass}

\subsection{Diverging effective masses for the $[001], [111]$, and $[112]$ directions}
In the main text, Fig.~\ref{fig:effective_mass} shows how each growth direction features points in $\tilde{\delta}_\epsilon$ where the effective mass $1/m^*$ diverges. 
In this Appendix, we show additional results for the $[001], [111]$, and $[112]$ directions not shown in the main text. 
While the behavior for the different growth directions is qualitatively the same as compared with the $[110]$ direction, we note that the excitation gaps in the different directions generally differ.
This yields quantitatively different convergence of the perturbative series as visible for the $[111]$ direction for small $|\tilde{\delta}_\epsilon|$.

\begin{figure*}
    \centering
    \includegraphics[width=\linewidth]{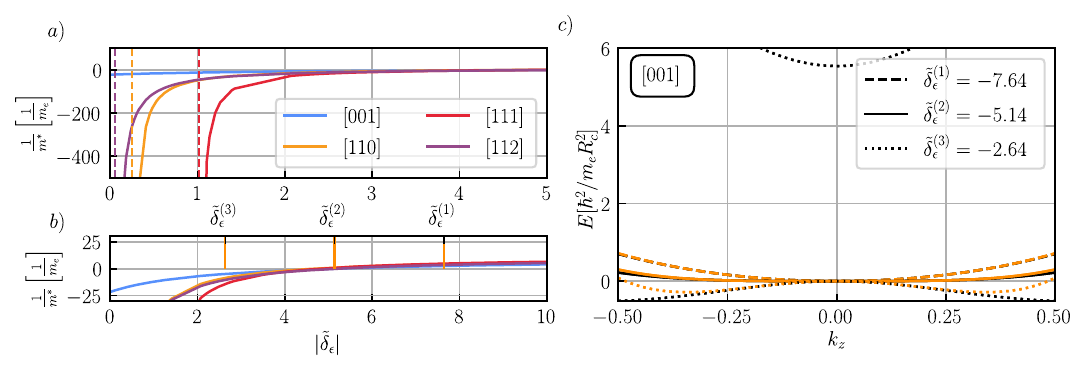}
    \caption{Inverse effective electron mass and bandstructures in the vicinity of the divergence points for the $[001]$ direction. In c), black lines indicate exact diagonalization, orange lines indicate energies of the perturbative expansion.}
    \label{fig:eff_mass_001}
\end{figure*}

\begin{figure*}
    \centering
    \includegraphics[width=\linewidth]{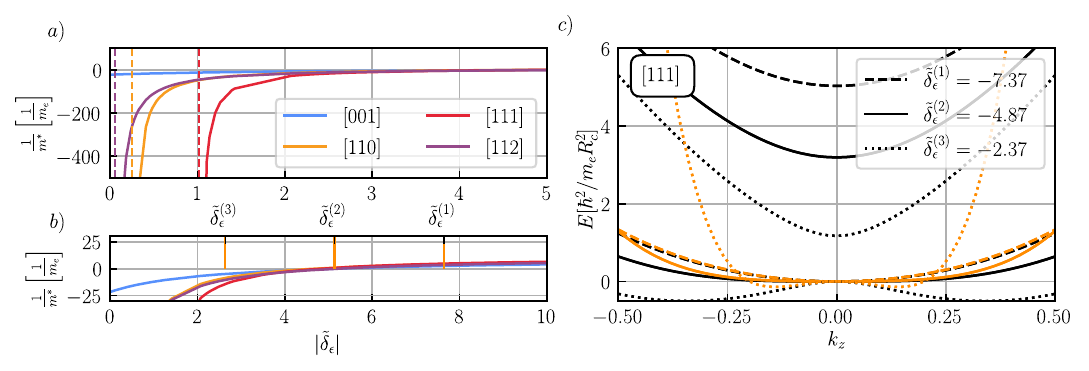}
    \caption{Inverse effective electron mass and bandstructures in the vicinity of the divergence points for the $[111]$ direction. In c), black lines indicate exact diagonalization, orange lines indicate energies of the perturbative expansion.}
    \label{fig:eff_mass_111}
\end{figure*}

\begin{figure*}
    \centering
    \includegraphics[width=\linewidth]{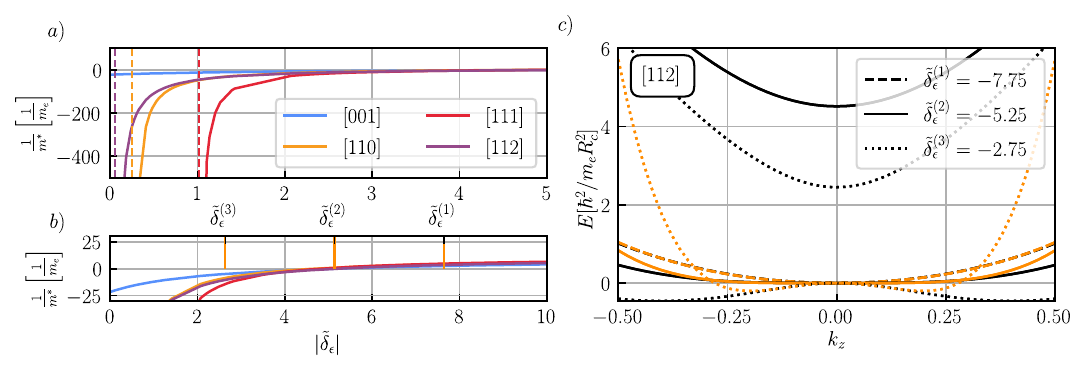}
    \caption{Inverse effective electron mass and bandstructures in the vicinity of the divergence points for the $[112]$ direction. In c), black lines indicate exact diagonalization, orange lines indicate energies of the perturbative expansion.}
    \label{fig:eff_mass_112}
\end{figure*}

\subsection{Effective mass dependency on electric and magnetic fields}

\begin{figure*}
    \centering
    \includegraphics[width=\linewidth]{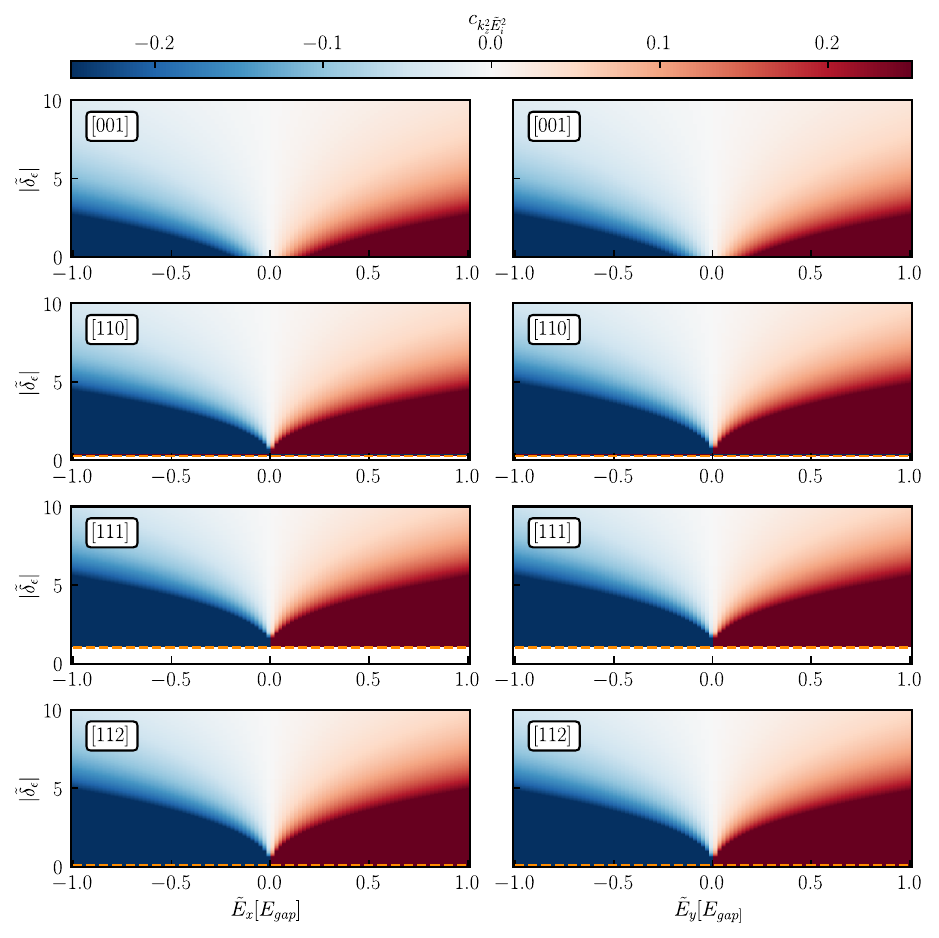}
    \caption{Perturbation theory coefficient $c_{k_z^2\tilde{E}_i^2}$ for electric fields applied in the $\hat{x}$ and $\hat{y}$ direction. Similar as for SOI, we choose $|eR_c\vec{E}|=E_{gap}$. We find that the response of the effective mass is largely independent of the direction of the fields.}
    \label{fig:eff_mass_square_e}
\end{figure*}

\begin{figure*}
    \centering
    \includegraphics[width=\linewidth]{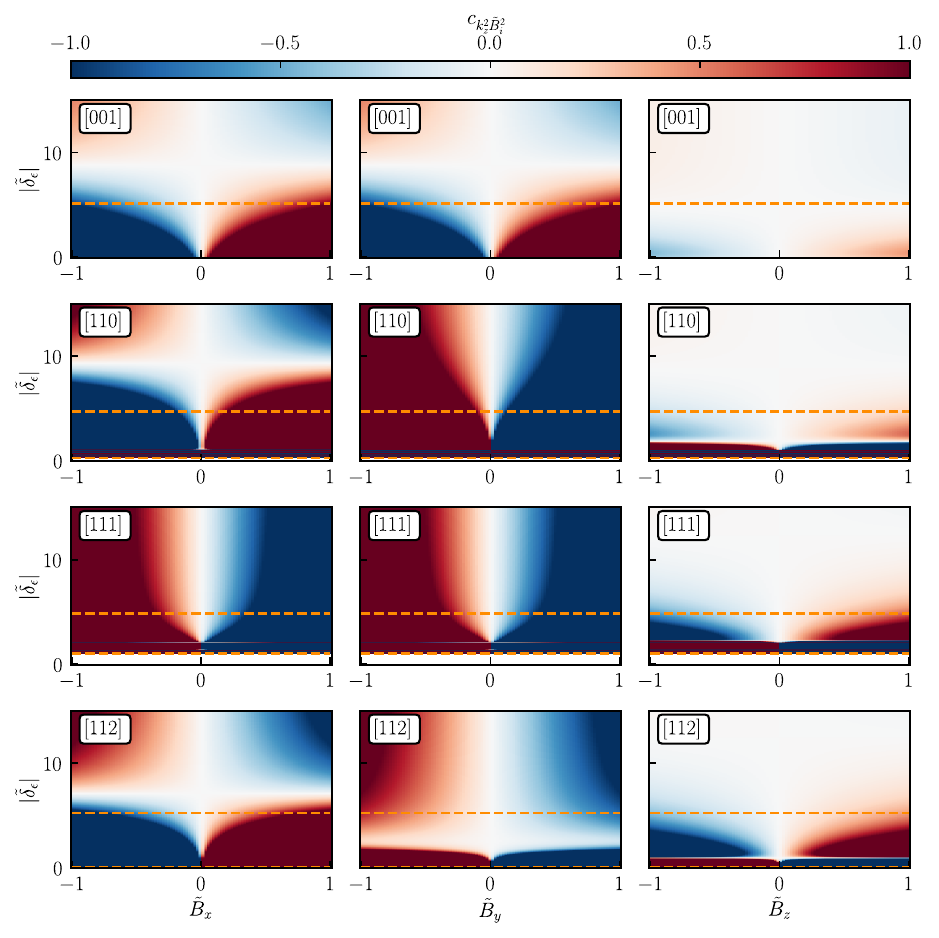}
    \caption{Perturbation theory coefficient $c_{k_Z^2\tilde{B}_i^2}$ depending of $|\tilde{\delta}_\epsilon|$ and $\tilde{B}_i$ for $i=x,y,z$. We find the correction to be anisotropic with respect to the direction of the applied magnetic field. Additionally, the coefficients change their sign in dependence on $|\tilde{\delta}_\epsilon|$. As indicated by orange dashed lines, the $\tilde{\delta}_\epsilon$ value where this happens differ from that of the orbital inversion and divergence of the bare effective mass.}
    \label{fig:eff_mass_square_b}
\end{figure*}

In eq.~\eqref{eq:eff_mass} in main text we note a dependency of the effective mass on externally applied electric and magnetic fields.
In this subsection we substantiate these claims by plotting the corresponding perturbative coefficients $c_{k_z^2\tilde{E}_i^2}$ and $c_{k_z^2\tilde{B}_i^2}$.
We plot the results for the electric field in Fig.~\ref{fig:eff_mass_square_e}, and magnetic fields in Fig.~\ref{fig:eff_mass_square_b}.
We find that the correction induced by electric fields depends in the coupling between ground and excited orbitals.
In contrast, the correction is independent of the direction of the applied electric field.
The magnetic field correction, shown in Fig.~\ref{fig:eff_mass_square_b}, shows to be anisotropic with the direction of the applied magnetic field.
Magnetic fields parallel to the wire generally lead to a smaller correction than for fields applied perpendicular to the wire.
Furthermore, Fig.~\ref{fig:eff_mass_square_b} shows that the correction coefficients change their sign depending on $\tilde{\delta}_\epsilon$.
As indicated by horizontal orange lines, these sign changes are however unrelated to both, the orbital inversions or the divergence of the bare effective mass.

\subsection{Additional results for in-situ effective mass tuning}

Beside the results for the $[110]$ direction shown in Sec.~\ref{fig:effective_mass_tuning} we here want to provide additional results for the $[001], [111]$, and $[112]$ growth directions.
The qualitative behavior is generally the same as presented in Fig.~\ref{fig:effective_mass_tuning}.

\begin{figure}
    \centering
    \includegraphics[width=0.9\linewidth]{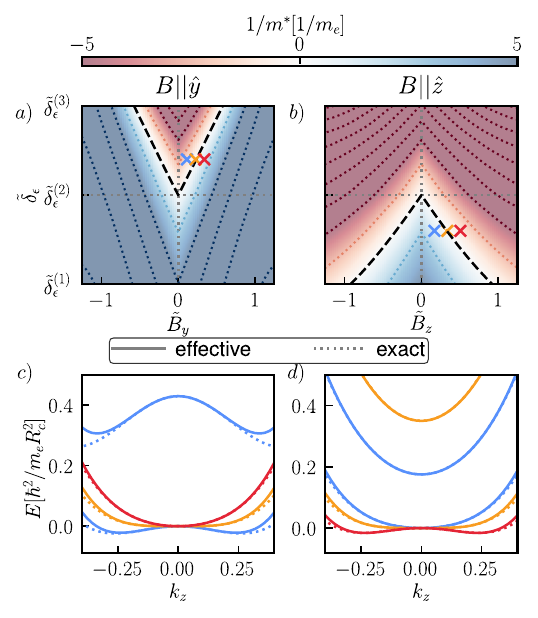}
    \caption{In-situ tuning of the effective mass with the magnetic field for the $[001]$ growth direction. The black dashed lines correspond to the solution to the parametric equation presented in eq~\eqref{eq:eff_mass_parametrization}.}
    \label{fig:eff_mass_tuning_001}
\end{figure}

\begin{figure}
    \centering
    \includegraphics[width=0.9\linewidth]{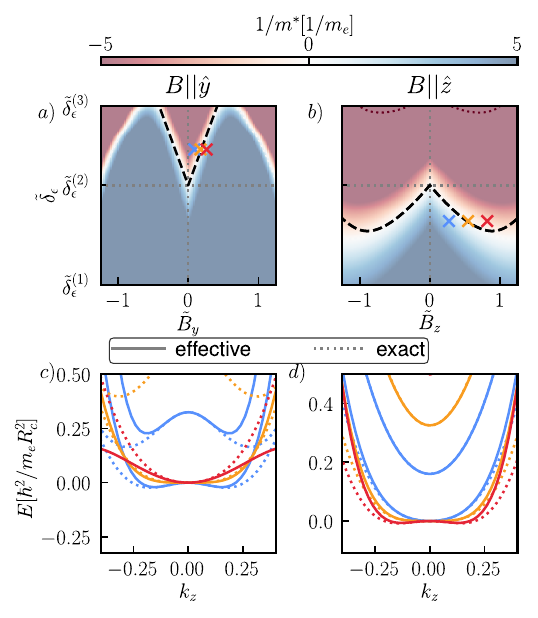}
    \caption{In-situ tuning of the effective mass with the magnetic field for the $[111]$ growth direction. The black dashed lines correspond to the solution to the parametric equation presented in eq~\eqref{eq:eff_mass_parametrization}.}
    \label{fig:eff_mass_tuning_111}
\end{figure}

\begin{figure}
    \centering
    \includegraphics[width=0.9\linewidth]{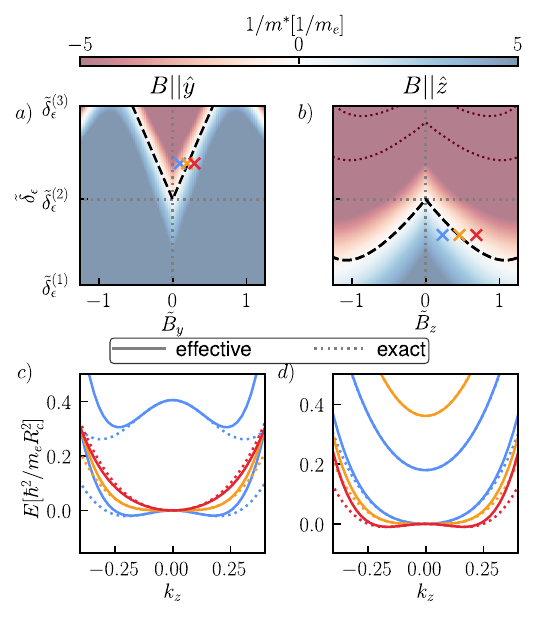}
    \caption{In-situ tuning of the effective mass with the magnetic field for the $[112]$ growth direction. The black dashed lines correspond to the solution to the parametric equation presented in eq~\eqref{eq:eff_mass_parametrization}.}
    \label{fig:eff_mass_tuning_112}
\end{figure}

\section{Additional results on spin-orbit interaction} \label{app:rashba}

\subsection{Linear and cubic spin-orbit interaction for $\vec{E}||\hat{y}$}

\begin{figure}
    \centering
    \includegraphics[width=\linewidth]{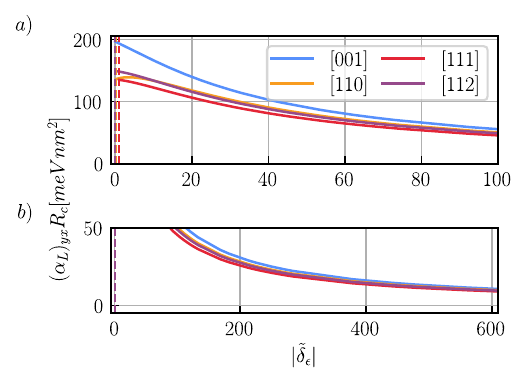}
    \caption{Linear spin-orbit interaction coefficients for electric fields parallel to the $\hat{y}$ direction.}
    \label{fig:linear_soi_y}
\end{figure}

\begin{figure}
    \centering
    \includegraphics[width=\linewidth]{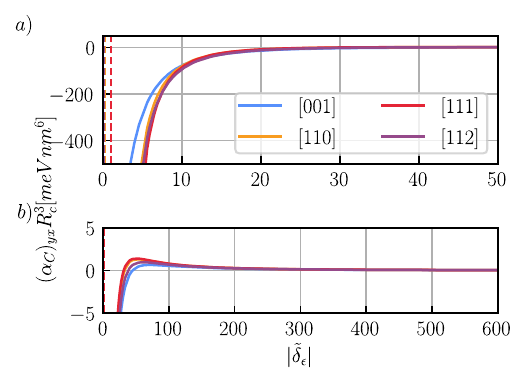}
    \caption{Cubic spin-orbit interaction coefficients for $\vec{E}||\hat{y}$.}
    \label{fig:cubic_soi_y}
\end{figure}

In the main text, we showed results for linear and cubic SOI for electric fields applied in the $\hat{x}$ direction.
We have performed the same analysis for electric fields in the $\hat{y}$ direction the results of which we collect here in Figs.~\ref{fig:linear_soi_y} and~\ref{fig:cubic_soi_y}.
Our results show that changing the alignment reverses the sign of the SOI.
We observe the same for cubic SOI in Fig.~\ref{fig:cubic_soi_y}.

\subsection{Higher-order correction of the linear spin-orbit interaction}

\begin{figure*}
    \centering
    \includegraphics[width=\linewidth]{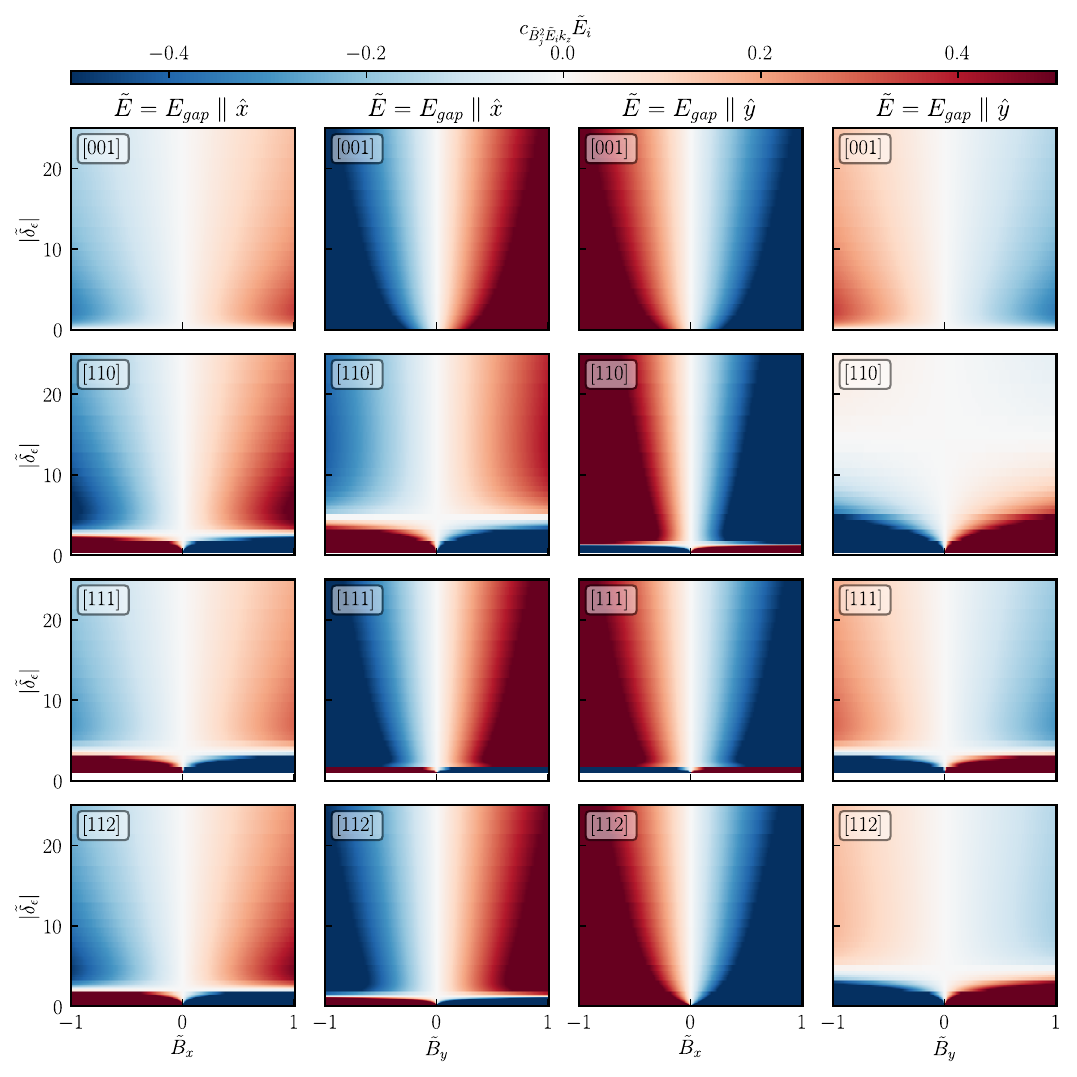}
    \caption{Corrections to the linear SOI coefficient through quadratic orders of magnetic field. We fix the electric field to the maximally possible value, $|eR_c\vec{E}|=E_{gap}$ and vary $\tilde{B}$. The correction is strongest in the vicinity of the orbital inversions due to the small excitation gap. Depending on the rescaled strain, it switches sign implying a rapid suppression with an increasing excitation gap. We note that the values of $|\tilde{\delta}_\epsilon|$ where the sign changes are different to the sign reversal points of the inverse effective mass.}
    \label{fig:soi_square_b}
\end{figure*}

In this section we collect the results on higher order corrections to the spin-orbit interaction presented in eq.~\eqref{eq:linear_rashba}, in Sec.~\ref{sec:rashba}.
As discussed in the main text, the linear Rashba coefficient couples to the magnetic field.
To linear order in $\vec{B}$, the SOI yields an energy shift for magnetic fields applied orthogonally to the electric field.
To quadratic order in $\vec{B}$, the magnetic field we find that the corresponding correction coefficient, $c_{\tilde{B}_i^2\tilde{E}_jk_z}$, generally changes sign depending on $\tilde{\delta}_\epsilon$.
We note that the corresponding values of $\tilde{\delta}_\epsilon$ where the sign changes depends on the growth direction and is different from the values where the effective mass (c.f. Sec.~\ref{Sec:effective_mass}) changes. 
Similarly, we find that, when the directions of magnetic and electric field are exchanged, the sign of the correction reverses.
Therefore it suffices to discuss the results for a single, fixed direction of the electric field, e.g., $\vec{E}||\hat{x}$.
Our results indicate that the correction is generally largest in the vicinity of the orbital inversions due to the magnitude of the SOI itself.
Due to SOI decreasing with increasing excitation gap, we find that the correction from the magnetic field decreases correspondingly.
We collect the results of our analysis in Fig.~\ref{fig:soi_square_b}.

\end{document}